\documentclass[a4paper,11pt]{article}
\usepackage{jheppub_mod}
\usepackage[T1]{fontenc}

\usepackage{hyperref}
\usepackage{graphicx}
\usepackage{amsmath,amssymb,mathtools,slashed}
\usepackage{booktabs,tabulary}
\usepackage{dsfont}
\usepackage{color}
\usepackage{multirow}
\usepackage{subcaption}
\usepackage{braket}
\usepackage{adjustbox}
\usepackage{footmisc}

\newcommand{\be}{\begin{equation}}
\newcommand{\ee}{\end{equation}}

\newcommand{\cO}{\mathcal{O}}

\newcommand{\GeV}{\,\text{GeV}}
\newcommand{\MeV}{\,\text{MeV}}

\renewcommand{\Im}{\text{Im}\,}
\renewcommand{\Re}{\text{Re}\,}

\newcommand{\chpt}{$\chi$PT}

\allowdisplaybreaks[1]

\title{Solutions of Roy equations for variable pion mass}

\author[a]{Gilberto Colangelo}

\affiliation[a]{
Albert Einstein Center for Fundamental Physics \\ Institute for Theoretical
Physics \\ University of Bern\\ Sidlerstrasse 5, 3012 Bern, Switzerland}

\emailAdd{gilberto@itp.unibe.ch}

\abstract{The Roy equations for the $S$ and $P$ waves are solved for pion
  masses below the physical value in the region between threshold and
  $s_A=0.84\GeV^2$. The analysis is based on the assumption that the
  phenomenological input above $s_A$ can be used essentially
  unchanged. Moreover, for $M_\pi \le M_{\pi,\mathrm{phys}}$, the chiral
  perturbation theory predictions for the two $S$-wave scattering lengths
  provide a very reliable additional input to the Roy equations. The
  outcome is an accurate description of the $S$ and $P$ partial waves as
  the light-quark masses decrease towards the chiral limit. As a test of
  the approach, we evaluate the first few threshold parameters of the $S$,
  $P$, and $D$ waves via Wanders' sum rules and find excellent agreement
  with chiral perturbation theory across the entire range $0 \le M_\pi \le
  M_{\pi,\mathrm{phys}}$. We further solve the Roy equations for pion masses
  above the physical value (up to $240\MeV$) and show that the resulting
  solutions agree well with existing lattice data at $M_\pi=239\MeV$. The
  present work paves the way for analyses of lattice data that fully
  incorporate the constraints of analyticity, unitarity, and crossing
  symmetry, and provides the basis for an accurate study of the dependence
  of the $\sigma$ and $\rho$ resonances on the light-quark masses, both of
  which will be addressed in future work. }

\begin{document} 

\maketitle

\section{Introduction}
\label{sec:intro}
The Roy equations~\cite{Roy:1971tc} incorporate the constraints of
analyticity and crossing symmetry for the $\pi \pi$ scattering amplitude
into a set of coupled equations relating different partial waves,
expressing the real parts as dispersive integrals over the imaginary
parts. Combining this with unitarity leads to a set of nonlinear, coupled
integral equations for the phase shifts. These can be solved numerically
(see~\cite{Basdevant:1972uu,Basdevant:1973ru,Pennington:1973xv,Pennington:1973hs}
for early attempts
and~\cite{Ananthanarayan:2000ht,Garcia-Martin:2011iqs,Caprini:2011ky} for
more recent ones) and yield an accurate determination of the $\pi \pi$
scattering amplitude below about $1\GeV$ as a function of the two $S$-wave
scattering lengths, which play the role of subtraction constants. Taking as
input the chiral perturbation theory
(\chpt)~\cite{Weinberg:1978kz,Gasser:1983yg} predictions for the
subtraction constants, which are known to
NNLO~\cite{Weinberg:1966kf,Gasser:1983yg,Bijnens:1995yn,Bijnens:1997vq},
reduces the uncertainties to the few-percent
level~\cite{Colangelo:2001df}. These \chpt~predictions have since been
confirmed experimentally to a comparable level of
precision~\cite{NA482:2005wht,NA482:2007xvj,NA482:2010dug}.

The combination of dispersion relations and \chpt\ represents a powerful
framework for obtaining very precise predictions for hadronic amplitudes in
regions where inelastic channels do not play a significant role. The $\pi
\pi$ scattering amplitude is a particularly successful example, but there
are many more (such as $\pi K$~\cite{Buettiker:2003pp,Pelaez:2018qny}, $\pi
N$~\cite{Ditsche:2012fv,Hoferichter:2015hva}, $\eta \to 3
\pi$~\cite{Kambor:1995yc,Anisovich:1996tx,Colangelo:2016jmc,Colangelo:2018jxw}
and others). On the one hand, chiral symmetry can fix the subtraction
constants, thereby pinning down the free parameters in dispersion
relations; on the other hand, this framework significantly extends the
region where precise predictions based on \chpt~can be made, without loss
of rigour, since it only relies on fundamental principles. However, this
approach requires phenomenological input at least in some energy
regions. Since such input can only be obtained for physical quark masses,
the ability to rigorously control the dependence on the light-quark masses
(\textit{i.e.} $m_u$ and $m_d$), as is possible within \chpt\ alone, appears
to be lost.

This is a significant drawback when seeking a connection with lattice QCD
calculations: predictive dispersive analyses were previously only
applicable to lattice simulations at the physical point. This paper
addresses this limitation and shows that the Roy equations can successfully
be solved for variable light quark masses below their physical values, all
the way down to the chiral limit. For this to be possible, one key
assumption is required: that the phenomenological input for the Roy
equations---namely the imaginary parts of the $S$ and $P$ waves above $\sim
1\GeV$, and for all other waves from threshold to infinity\footnote{While
this is technically the case, $D$ and higher waves below $1\GeV$ are
negligibly small and contribute very little to the
dispersive integrals. Thus, for these waves as well, the assumption
effectively concerns only the sensitivity of the region above $1\GeV$ to
the light quark masses.}---remains essentially unchanged as the light-quark
masses are lowered from their physical value to zero. This is a very
reasonable assumption, considering that the light quark masses are only a
few MeV and that the uncertainties in the phenomenological input are
typically much larger than a few percent. Moreover, as we show below, this
assumption can be tested directly.

One might wonder how this helps comparisons with lattice calculations,
given that physical light-quark masses are already difficult enough to
simulate, let alone even lighter ones. The point is that the low-mass
region serves mainly as a validation of the framework, which can then be
readily extended to pion masses above the physical value, connecting
directly with lattice calculations. The extension to heavier pion masses
remains valid up to a certain limit, because the assumption of pion-mass
independence of the phenomenological input will break down at some value of
the pion mass. Furthermore, the convergence of the chiral series for the
subtraction constants will also impose limitations on the region of pion
masses for which this approach remains valid. The aim of this first paper
on this topic is not to thoroughly investigate these questions, but rather
to provide a proof of principle that such an approach can succeed.

Providing guidance for understanding the pion-mass dependence of quantities
which are beyond the chiral regime would have a significant impact on the
analysis of lattice data. Quantities in this category include the partial
waves of $\pi \pi$ scattering at energies between $0.5$ and $1\GeV$, which
can be computed directly on the lattice, as well as the properties of the
lightest resonances, notably the $\sigma$ and the $\rho$. Calculating pole
positions in the complex plane requires analytic continuation, for which
dispersion relations provide the only model-independent and rigorous
framework~\cite{Caprini:2005zr,Garcia-Martin:2011nna,Moussallam:2011zg},
that is however problematic for unphysical pion masses, as discussed
above. The analysis presented here offers a solution to this problem
(recent, alternative approaches which rely on the same basic principles,
but differ in the implementation, can be found in
Refs.~\cite{Cao:2023ntr,Rodas:2023nec} and will be discussed in more detail
below). Early studies of this topic were presented in
Ref.~\cite{Hanhart:2008mx}, and more recently in
Refs.~\cite{Molina:2020qpw,Niehus:2020gmf,Pelaez:2026fmu}. These works rely
on a unitarization of \chpt, specifically the inverse amplitude method
(IAM), to describe the quark-mass dependence of observables outside the
region of convergence of \chpt. There is a considerable body of literature
on this approach and its applications to phenomenology (see, for example
Refs.~\cite{Truong:1988zp,Dobado:1989qm,Dobado:1992ha,Dobado:1996ps,
  Guerrero:1998ei,GomezNicola:2001as,Pelaez:2004xp,GomezNicola:2007qj}). While
the IAM handles the right-hand cut rigorously, it accounts for the
left-hand cut only through the values of the LECs, rather than treating
both cuts on the same footing. The approaches in
Refs.~\cite{Cao:2023ntr,Rodas:2023nec} as well as the present one go beyond
this approximation. Here, in particular, we work with exact numerical
solutions of the Roy equations, confining the approximations to the
phenomenological input, where they are unavoidable. A dedicated study of
the lowest resonances of QCD based on the present work will appear in a
subsequent paper~\cite{Colangelo-Zwicky}. This is also motivated by the
long-standing idea that the $\sigma$ meson might be interpreted as the
dilaton of spontaneously broken scale invariance in
QCD~\cite{Ellis:1970yd,Crewther:1970gx,
  Crewther:2013vea,Golterman:2016lsd,Zwicky:2023krx} (see
Ref.~\cite{Pelaez:2015qba} for a review about the $\sigma$ meson).

The rest of the paper is organized as follows: Sect.~\ref{sec:royeq}
provides a concise summary of the Roy equations and their numerical
solutions for physical quark masses. In Sect.~\ref{sec:roympi} the
framework for solving the Roy equations for quark masses below their
physical value is spelled out, including all the necessary
approximations. The numerical solutions are presented and discussed in
Sect.~\ref{sec:numsol}, along with a detailed uncertainty
estimate. Sect.~\ref{sec:disp-chpt} tests the chiral behaviour of the
solutions by evaluating threshold parameters via Wanders' sum rules, while
Sect.~\ref{sec:heavierMpi} discusses the extension of the present framework
to pion masses heavier than the physical value and offers a comparison to
existing lattice data. Conclusions and an outlook are presented in
Sect.~\ref{sec:concl}.

\section{Numerical solutions of the Roy equations for physical quark masses}
\label{sec:royeq}
The Roy equations for the $\pi\pi$ scattering amplitude incorporate the
constraints of analyticity, unitarity and (partially) crossing symmetry,
and have the following form~\cite{Roy:1971tc}:
\begin{equation}
\label{eq:Roy1}
t_\ell^I(s)= k_\ell^I(s)+\sum_{I'=0}^2\sum_{\ell'=0}^\infty \int_{4M_\pi^2}^\infty
ds'\,K_{\ell\ell'}^{I I'}(s,s')\,\Im t_{\ell'}^{I'}(s')\, ,
\end{equation}
where $I$ and $\ell$ indicate isospin and angular
momentum, respectively, and $k^I_\ell(s)$ is the subtraction polynomial,
which can be expressed in terms of the two $S$-wave scattering lengths,
$a_0^I$ as follows: 
\begin{align} 
\label{eq:kIl}
k^I_\ell(s) &= a_0^I \,\delta_\ell^0 +\frac{s-4M_\pi^2}{4M_\pi^2}\,
(2a_0^0-5a_0^2)\,\left(\frac{1}{3}\,\delta_0^I\,\delta_\ell^0
+\frac{1}{18}\,\delta_1^I\,
\delta_\ell^1-\frac{1}{6}\,\delta_2^I\,\delta^0_\ell\right) \, ,
\end{align}
and is non-zero only for $S$- and $P$-waves. The kernels
$K_{\ell\ell'}^{I I'}(s,s')$ are known and contain a diagonal, singular
Cauchy kernel that generates the right-hand cut in the partial-wave
amplitudes, as well as a logarithmically singular piece that accounts for
the left-hand cut. In concrete applications one typically considers only
$S$ and $P$ waves and solves the equations for them up to a certain energy
of $\sqrt{s_u}\sim\cO(1\GeV)$. Between this and $\sqrt{s_2}\sim \cO(2\GeV)$
one takes a phenomenological input for the imaginary parts of the same
waves. This contribution can be calculated once and for all and will be
denoted by $f^I_\ell(s)$. The contributions from yet higher energies and of
higher waves are estimated from phenomenology and collected into the
driving terms $d^I_\ell(s)$. With this splitting of different
energy contributions, the Roy equations can be written as follows:
\begin{align}
\label{eq:Roy2}
\Re t_\ell^I(s)= k_\ell^I(s)&+\int_{4M_\pi^2}^{s_u} ds'\,K_{\ell 0}^{I
  0}(s,s')\,\Im t_0^0(s')
+\int_{4M_\pi^2}^{s_u} ds'\,K_{\ell 1}^{I 1}(s,s')\,\Im t_1^1(s') \nonumber
\\
&+\int_{4M_\pi^2}^{s_u} ds'\,K_{\ell 0}^{I 2}(s,s')\,\Im
t_0^2(s')+f^I_\ell(s)+d^I_\ell(s) =: \text{Roy}^I_\ell(s) \; .
\end{align}

The first comprehensive modern analysis of the Roy equations was carried
out in Ref.~\cite{Ananthanarayan:2000ht}, which
reviewed many theoretical aspects and experimental inputs of the
equations. In that paper the Roy equations were solved below
$\sqrt{s_0}=0.8\GeV$ (setting $s_u=s_0$ in~\eqref{eq:Roy2}), a value chosen
to lie between the energies where the phase shifts $\delta_1^1$ and
$\delta_0^0$ cross $\pi/2$. With this choice the solution is guaranteed to
be unique, for each pair of $S$-wave scattering lengths $a_0^0$ and $a^2_0$
that lie within the universal band (UB). In a subsequent
paper~\cite{Colangelo:2001df} theoretical constraints about the two
subtraction constants coming from chiral symmetry have been used, which has
led to much tighter predictions. In the meantime the theory predictions for
the two $S$-wave scattering lengths have been confirmed experimentally,
with a similar level of
precision~\cite{NA482:2005wht,NA482:2007xvj,NA482:2010dug}, establishing
the analysis in~\cite{Colangelo:2001df} as the benchmark for the
phenomenology of the $\pi \pi$ interaction.

A further analysis of the Roy equations was performed about ten years
later~\cite{Caprini:2011ky} and improved on a number of aspects:
\begin{enumerate}
\item the Roy equations for the $S$ and $P$ waves were solved up to
$s_1=68 M_{\pi,\mathrm{phys}}^2$ ($\sqrt{s_1}\simeq1.15\GeV$), the upper limit of
  validity of the equations  (setting $s_u=s_1$ in~\eqref{eq:Roy2});  
\item the Roy equations were solved also for $D$ and $F$
  waves\footnote{Though less relevant, the input for $G$ and $H$ waves has
  also been used and checked for consistency with the Roy equations.}
  (all of them up to $s_1$);
\item sum rules imposing full crossing symmetry were implemented;
\item continuous transitions in each channel between the description in
  terms of partial waves (below $\sqrt{s_2}=1.6\GeV$) and the one in terms
  of a Regge parametrization (above $s_2$) were imposed.
\end{enumerate}
Instead of solving the Roy equations for all these waves and imposing the
sum rules and continuity conditions at the same time, all this was
accomplished with an iterative procedure which, however, converged after
very few iterations. Another important difference with respect to the
analysis in~\cite{Ananthanarayan:2000ht} is that by raising the matching
point to $s_1$, where the $\delta_0^0$ phase has reached values above
$3\pi/2$, the system has gained three degrees of freedom in the
multiplicity of solutions. This means that in addition to the phase values
at the matching point $s_1$ one needs to input three more parameters to
uniquely pin down the solution. In~\cite{Caprini:2011ky} these parameters
were $\delta_0^0(s_0)$, $\delta_1^1(s_0)$ and $\delta_0^0(4 M_K^2)$, and
the whole set of phase input values was:
\begin{align}
\delta_0^0(s_0)&=(82.3\stackrel{+10}{_{-4}})^\circ, \quad \delta_0^0(4 M_K^2)=(185\pm
10)^\circ, \quad \delta_0^0(s_1)=(260 \pm 10)^\circ, \nonumber \\
\delta_1^1(s_0)&=(108.9\pm 2)^\circ, \hskip 4.65cm \delta_1^1(s_1)=(166.5\pm
2)^\circ.
\end{align}
In the present analysis we rely on the solutions obtained
in~\cite{Caprini:2011ky} and on the corresponding input parameters, with
one exception: the uncertainty in the $S0$ wave is reverted to
$\delta_0^0(s_0)=(82.3\pm 3.4)^\circ$, as in~\cite{Ananthanarayan:2000ht},
since the error stretching adopted in~\cite{Caprini:2011ky} no longer
appears justified. One could in principle consider the analysis
in~\cite{Colangelo:2018mtw}, which has shown that the $P$ wave phases are
now known with even smaller uncertainties, $\delta_1^1(s_0)=(110.4\pm
0.3)^\circ$ and $\delta_1^1(s_1)=(165.7\pm0.5)^\circ$ (see
also~\cite{Leplumey:2025kvv}). On the other hand, since these values are
fully compatible with those adopted in~\cite{Caprini:2011ky}, 
and the uncertainty reduction does not play an important role in the
present analysis, we keep the central values and uncertainties in
Ref.~\cite{Caprini:2011ky}, for simplicity. Finally, since $s_1$ is above the
inelastic threshold, one also needs to input the elasticity for all
waves. This part of the input is adopted directly from~\cite{Caprini:2011ky}.

\section{Solving the Roy equations for $M_\pi < M_{\pi,\mathrm{phys}}$}
\label{sec:roympi}
The goal of the present analysis is to find solutions of the Roy equations
in a world where the light quark masses are below their physical
values. Between threshold and the matching point $s_u$ the necessary
modification is generated self-consistently by the Roy equations
themselves. The changes we have to explicitly implement concern the
position of the threshold (in physical units) and the value of the two
subtraction constants, $a^{0,2}_0$, whose dependence on the pion mass is
predicted by \chpt, and explicitly known up to $\cO(p^6)$. In principle we
would have to adapt the input for the $S$ and $P$ waves above $s_1$ to
lower values of the quark masses, but as we anticipated in the
introduction, it is reasonable to assume that this would barely change,
very likely in the few-percent range. In view of the much larger
uncertainties affecting this input, we will neglect such a change
altogether. Other inputs to the Roy equations are $D$ and higher
waves. While it is true that we need to provide them as input already from
threshold, the kinematic suppression of higher partial waves ($q^{2
  \ell+1}$ for imaginary parts) renders any detailed knowledge of their
variation near threshold---including the shift of the threshold position
itself---unnecessary. As is well known, $D$ and higher waves contribute
significantly to the Roy equations for the $S$ and $P$ waves (via the
driving terms) only through the resonances appearing in these
channels. These have masses well above $1\GeV$ and it is reasonable to
assume that also their properties will change within uncertainties if the
light quark masses are lowered towards the chiral limit.

\begin{figure}[t]
  \centering
  \includegraphics[width=7.5cm]{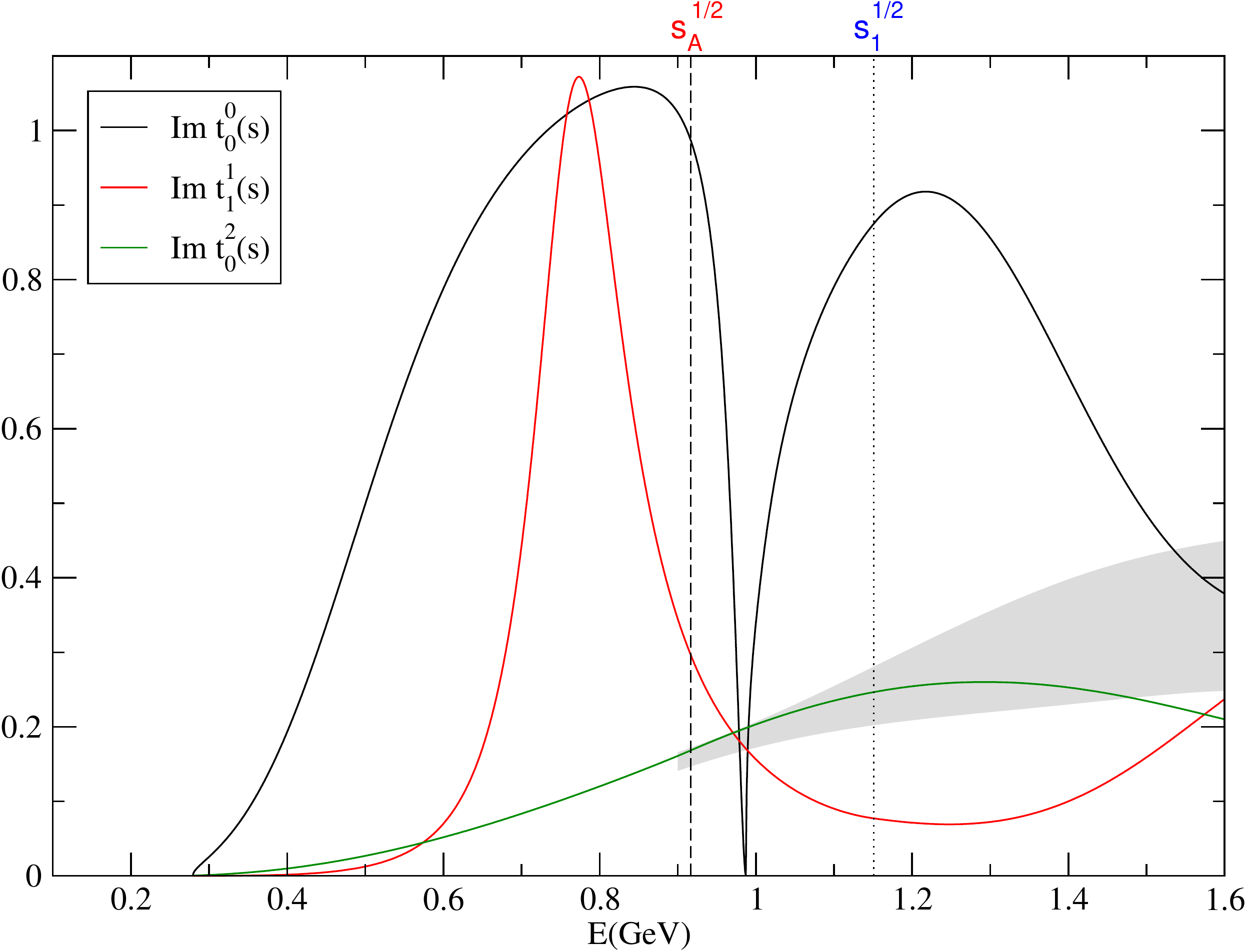}
  \includegraphics[width=7.5cm]{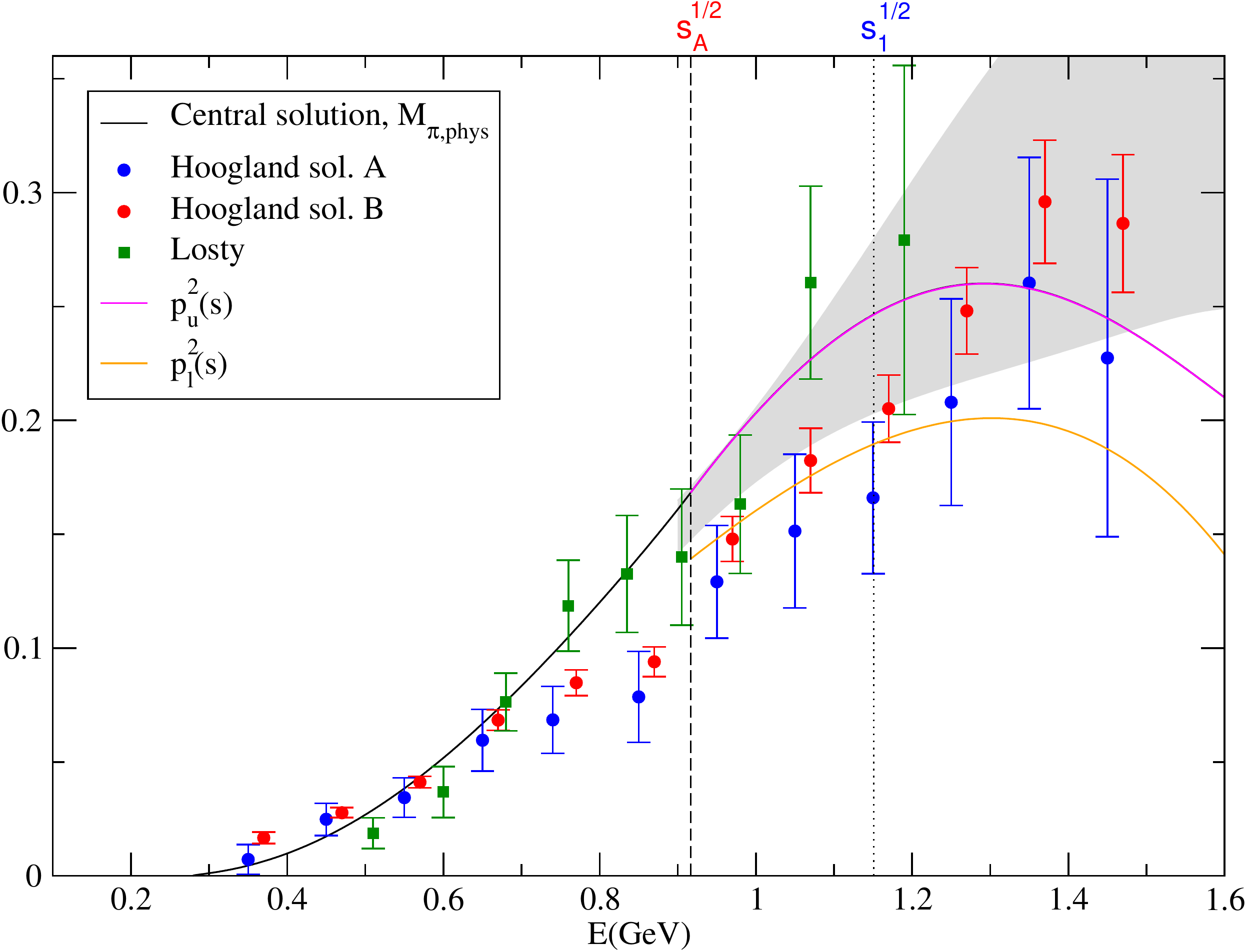}
  \caption{\label{fig:imtI_central} (Left panel) Central solutions with
    central values of input parameters as in~\cite{Caprini:2011ky}. (Right
    panel) Data for $\Im t^2_0(s)$ together with the two curves among which
    we interpolate to describe possible behaviours of this partial wave
    above $s_A$. The shape of the present input above $s_A$ for the $S2$
    wave differs from the one in~\cite{Caprini:2011ky} which is shown as
    a grey band in both panels.}
\end{figure}

With a matching point at $1.15\GeV$ another relevant input is the set of
inelasticities. While those for the $P$ and $S2$ waves are small and start
very smoothly, the one for the $S0$ wave opens up very rapidly, with a
square root behaviour and a large coefficient at the $K\bar{K}$
threshold. This is related to the presence of the $f_0(980)$ resonance
which is very close to this threshold and couples both to $\pi \pi$ and
$K\bar{K}$. This resonance induces rapid changes in the $t_0^0$ wave, which
require an {\em ad hoc} parametrization to be described correctly, like the
Flatt\'e parametrization adopted in~\cite{Caprini:2011ky}. The parameters
of the $f_0(980)$ and the corresponding inelasticity were taken as input in
that work, and only the smooth background was left to the Roy equations to
determine. Expecting this background to remain essentially unchanged as the
light-quark masses decrease, we take the entire region from just below the  
 $K\bar{K}$ threshold upward as input, thereby lowering the matching point
to $s_A=0.84\GeV^2$. Figure~\ref{fig:imtI_central} shows the Roy solutions
corresponding to the central values of the input parameters as
in~\cite{Caprini:2011ky}. Only for the $S2$ wave is a somewhat different
input adopted, as also shown in the same figure, for reasons that are
explained below. 

As shown in~\cite{Ananthanarayan:2000ht}, the requirement that in this wave
there is no cusp at the matching point imposes a correlation between the
behaviour of this wave at the matching point and the $a^2_0$ scattering
length. By continuity, a change of the $S2$ partial wave at $s_A$ implies a
change also in a region above that energy. Since data show a mild
energy dependence between $s_A$ and $s_2=(1.6\GeV)^2$, the partial wave in
this region will be described by a low-degree polynomial, whose parameters
are fixed by requiring continuity of the function and its derivative at
$s_A$ and by the value at $s_2$ which is one of the inputs. To simplify the
analysis further, a correlation between the values at $s_A$ and $s_2$ has
been adopted: the behaviour of this wave in this region is described as a
linear interpolation between two smooth curves of the form:
\begin{equation} \Im
t^2_0(s) = x^2_0 p^2_u(s)+(1-x^2_0)p^2_l(s) \qquad s_A < s < s_2
\end{equation}
where
$p^2_{u,l}(s)$ are fourth-order polynomials in $s$, and are shown in the
right panel of Fig.~\ref{fig:imtI_central}, together with three datasets
from Refs.~\cite{Losty:1973et,Hoogland:1977kt}. The two curves are designed
to broadly go through the central portion of the region covered by the data
sets (and appear to do so better than the central value and uncertainty
band adopted in~\cite{Caprini:2011ky}, which are also shown in the figure,
but the change has barely any consequence on the solutions). The only free
parameter in this wave is $x^2_0$, which is determined implicitly by
$a^2_0$ when solving the Roy equations. The Roy-equation solution 
for physical pion mass and central values of the scattering lengths selects
the value of $x^2_0=1$, as also shown in
Fig.~\ref{fig:imtI_central}\footnote{The fact that the chiral value of
$a^2_0$ requires $\Im t^2_0(s_A)$ to be roughly at the upper end of the
allowed experimental range corresponds to the fact that the tiny \chpt\ ellipse
in the $(a_0^0,a_0^2)$ plane, see Fig.~1 in~\cite{Colangelo:2001df}, lies
at the lower end of the UB. In the present setting $x^2_0$ is exactly 1
because $p^2_u(s)$ is defined to correspond to the central curve, for the
chiral input for the scattering lengths.}. In general, however, when
solving the Roy equations we leave the parameter $x^2_0$ to be completely
free and do not constrain it to be between 0 and 1, so that curves above
the upper one or below the lower one are also allowed.

\enlargethispage{20pt}
On the more technical side, one novelty of the present analysis with
respect to previous ones~\cite{Ananthanarayan:2000ht,Caprini:2011ky}
concerns the parametrization of the phase adopted to solve the Roy
equations. While in previous analyses we relied on a Schenk parametrization
for the phase---essentially an extended threshold expansion---in the
present work a Chebyshev parametrization of the following form is introduced:
\begin{align}
\delta^I_\ell(s)&=\arctan \left[ \sigma(s) F^I(s) \frac{4 M_\pi^2 -
    s^I_R}{s-s^I_R} \right] \qquad I=0,1, \quad \ell=I \nonumber \\*
\delta^2_0(s)&= \sigma(s) F^2(s)  \nonumber \\*
F^{0,2}(s)&=a^{0,2}_\ell \delta_{\ell 0}+y(s) \sum_{k=0}^8 c^{0,2}_k
C_k(x(s)) \, ,
\quad F^1(s)=\frac{s-4 M_\pi^2}{4 M_\pi^2} \sum_{k=0}^8 c^{1}_k
C_k(x(s))
\label{eq:ChebI}
\end{align}
with $y(s)=(s-4 M_\pi^2)/(s_A-4 M_\pi^2)$ and $x(s)=2 y(s)-1$,
$\sigma(s)= \sqrt{1-4M_\pi^2/s}$ and $C_k(x)$
the Chebyshev polynomials of degree $k$. Since there is no resonance in the
$S2$ wave neither a pole parameter, nor the $\arctan$ is included in its
parametrization. Of the nine free parameters in the
Chebyshev parametrization of each wave, two are used to ensure continuity
of the imaginary part and of its first derivative at $s_A$. Overall, we
therefore have eight free parameters for the $S0$ and $P$ waves and seven
for the $S2$ wave. The change from a Schenk to a Chebyshev parametrization
made the numerical search for solutions of the Roy equations significantly
more efficient and allowed us to achieve solutions of better quality for
the same input parameters. 

\subsection{Defining the framework}
In principle, the setting described above completely defines the framework
in which to look for solutions of the Roy equations for quark masses lower
than physical. The quality of the solutions has been measured by the
following sum of squares: 
\begin{equation}
Q^2=\sum_{I=0}^2\sum_{k=1}^{24} [\Re
  t^I_\ell(s_k)-\text{Roy}^I_\ell(s_k)]^2 \; ,
\end{equation}
(with $\ell=0$ for $I=0,2$ and $\ell=1$ for $I=1$)
where $\Re t^I_\ell(s_k)=\sin(2 \delta^I_\ell(s_k))/(2 \sigma(s_k))$
is the real part calculated directly from the phase shift, and
$\text{Roy}^I_\ell(s_k)$ is the right-hand side of the Roy
equations~\eqref{eq:Roy2}, and $s_k$ are 24 suitably chosen energy values
between threshold and $s_A$. Attempts to find such solutions showed,
however that, as the pion mass was lowered, the quality quickly degraded,
{\em i.e.} the value of $Q^2$ quickly increased to a level where it could
hardly be called a solution anymore. The conclusion after such attempts was
that leaving the input above $s_A$ completely unchanged (modulo the degree
of freedom in the $S2$ wave represented by the parameter $x^2_0$) when 
lowering the pion mass is not compatible with the constraints imposed by
the Roy equations. To find solutions it is necessary to introduce more
degrees of freedom.

We have therefore adopted the following framework, which proves to be very
effective. In addition to $x^2_0$, which determines $\Im t_0^2(s_A)$, a
parameter controlling a relative shift in $\Im t^0_0(s_A)$ is introduced:  
\begin{equation}
\Im t^0_0(s_A)=\Im \overline{t^0_0(s_A)}(1+x_{It0})
\end{equation}
where $\Im \overline{t^0_0(s_A)}=0.987$ is the central
value. Moreover, two parameters are introduced, one for each of the
two $S$ waves, allowing for a relative shift in the derivative of the
imaginary parts at $s_A$:
\begin{equation}
\frac{d}{d s} \Im t^{0,2}_0(s_A)=\frac{d}{d s} \Im
\overline{t^{0,2}_0(s_A)}(1+x_{dIt{0,2}}) \;. 
\end{equation}
For continuity, these shifts in the input affect also the region above
$s_A$, but are driven smoothly to zero at $s_1$, so that above that energy
the input coincides with the one used in~\cite{Caprini:2011ky}. The
approach to solving the Roy equations works then as follows: for each set
of these four ``hyperparameters'' ($x_{It0}$, $x_{dIt{0}}$, $x^2_0$, $x_{dIt{2}}$)
$Q^2$ is minimized with respect to the Chebyshev coefficients of the
partial waves. A second, outer minimization is then performed to find the
minimum of the minimized $Q^2$ in the three-dimensional parameter space
spanned by ($x_{dIt{0}}$, $x^2_0$, $x_{dIt{2}}$). The parameter $x_{It0}$
is excluded from this outer minimization procedure because, in the range of
interest, $Q^2$ depends only very mildly on it: in other words, the space
of solutions exhibits an approximate degeneracy along this direction. Its
value must therefore be fixed by external phenomenological input. This is
not surprising, since fixing this parameter directly constrains the value
of the phase $\delta^0_0$ at $\sqrt{s_0}=0.8\GeV$, which is one of the
standard external inputs. In~\cite{Ananthanarayan:2000ht} the range
$\delta^0_0(s_0)=(82.3 \pm 3.4)^\circ$ was adopted and we retain this
choice here. This translates into the range 
\begin{equation}
x_{It0}(M_{\pi,\mathrm{phys}})=(0.0\pm2.93)\cdot 10^{-2} 
\end{equation}

\begin{figure}[t]
  \centering
  \includegraphics[width=7.5cm]{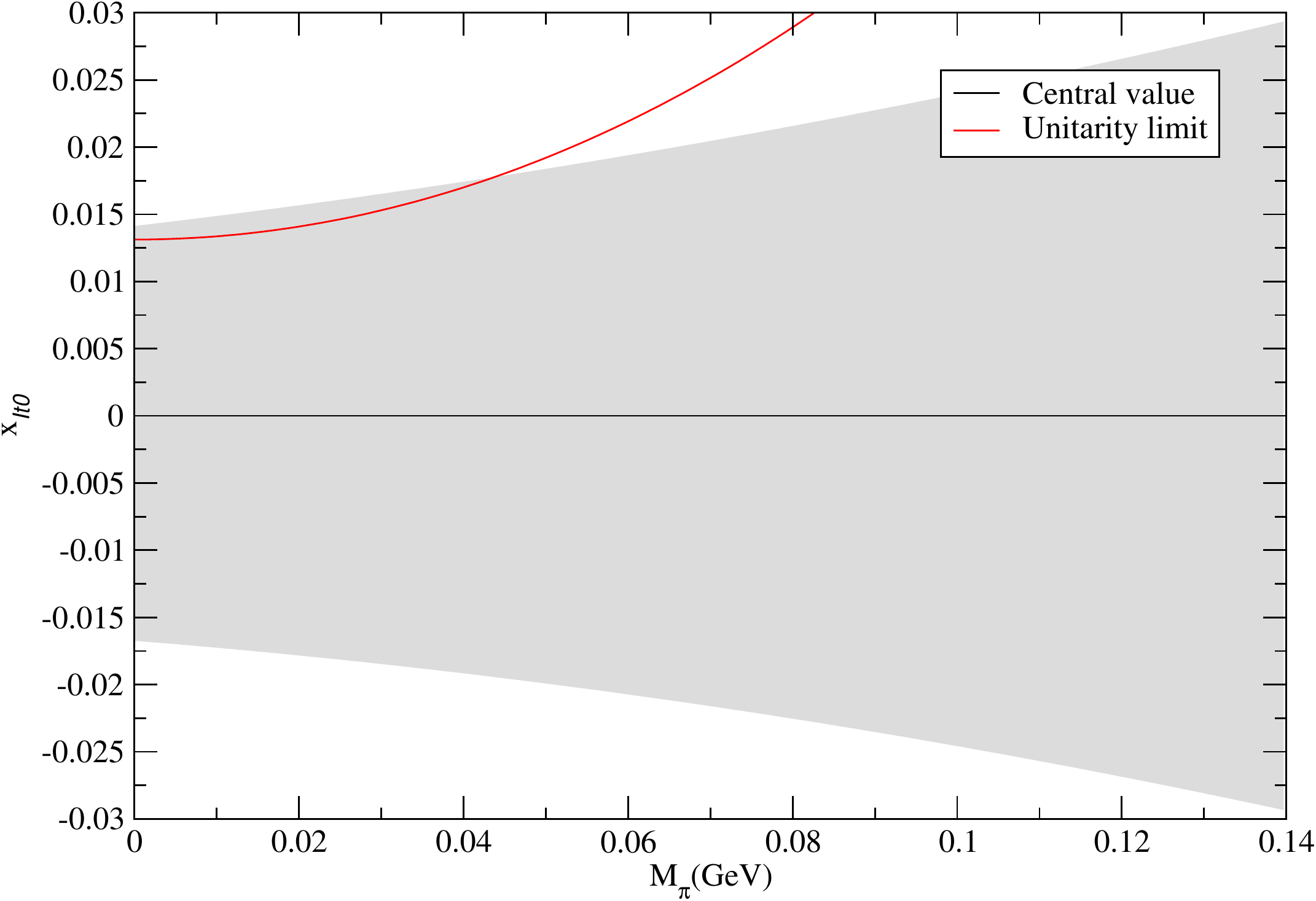}
  \includegraphics[width=7.5cm]{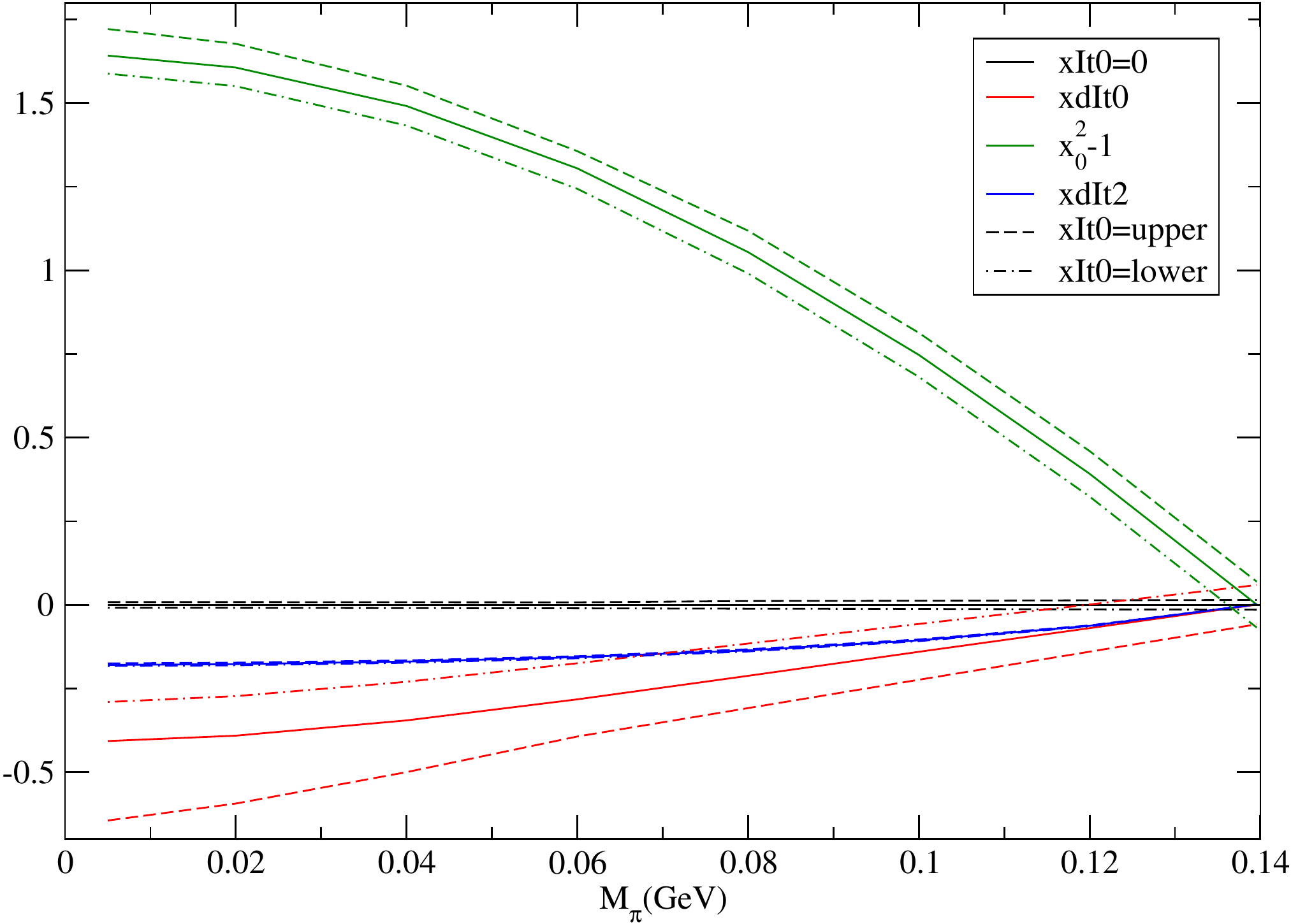}
  \caption{\label{fig:hyperp} Left: range of $x_{It0}$ for variable
    pion mass and unitarity upper limit for this parameter. Right:
    $M_\pi$ dependence and uncertainty range of the four hyperparameters. } 
\end{figure}

A good test of the soundness of this approach is to introduce these
hyperparameters for the physical pion mass and examine whether an even
lower $Q^2$ can be obtained. Since $x_{dIt{0}}$ and $x_{dIt{2}}$ represent
relative deviations from the central values of the adopted phenomenological
input, finding a lower $Q^2$ at non-zero values of these parameters (for
the central solution $x_{It0}=0$) would indicate that the phenomenological
input itself could be refined to better satisfy the Roy equations. It is
also instructive to analyse the shape of the $Q^2$ as a function of
$x_{It0}$. The phenomenological input passes this test convincingly: for
$x_{It0}=0$ the $Q^2$ minimum is found at $x_{dIt{0}}=5.9\cdot 10^{-4}$ and
$x_{dIt{2}}=1.3\cdot 10^{-3}$, while a scan in $x_{It0}$ around zero yields
a minimum at $x_{It0}=-2.1\cdot 10^{-3}$---all remarkably close to
zero. Furthermore, the additional freedom does not induce any shift in
$x^2_0$ which remains fixed at 1. These values are shown in
Fig.~\ref{fig:hyperp}, which also shows that the minima smoothly drift away
from zero, the central values at $M_{\pi,\mathrm{phys}}$, as the
pion mass decreases towards the chiral limit. This confirms the need to
adjust the phenomenological input at lower pion masses. At the same
time, because the variations in these parameters remain smooth and small,
the calculation supports our core assumption that the input above $s_A$
depends only mildly on the pion mass, or at least shows that it is fully
compatible with the constraints imposed by the Roy equations.

One may wonder at this point why no additional degrees of freedom have been
introduced for the $P$-wave. The main reason is practical:
when the quality of the solution begins to degrade, the dominant
contribution to the growth of $Q^2$ comes from the two $S$
waves. It was therefore natural to improve the solutions by
providing additional flexibility in these two waves. Of course, the input of
the $P$ wave is also subject to uncertainties; its variations are taken
into account as well and will be discussed below.

Within this extended framework, while the quality of the solutions still
decreases as the chiral limit is approached, the residuals remain very
small. The Roy equations have been solved for pion masses of $M_\pi=120$,
100, 80, 60, 40, 20 and $5\MeV$; $Q^2$ increases from
$\sim 10^{-10}$ at the physical point to $\sim 2 \cdot 10^{-9}$ at $80\MeV$ and
$1.2\cdot 10^{-8}$ at $5\MeV$. In terms of the root-mean-square (RMS) difference
between left- and right-hand side of the Roy equations
($\text{RMS}=\sqrt{Q^2/72}$ across the 24 points and three partial waves),
this corresponds to an increase from $10^{-6}$ at the physical point to
$1.3\cdot 10^{-5}$ at $5\MeV$, which still indicates a solution
of excellent quality.

When lowering the quark masses a pertinent question concerns the treatment
of uncertainties, which are well defined phenomenologically at the physical
point, but become somewhat arbitrary at lower pion masses. Our strategy
is as follows: even for lighter pions, the size
of the uncertainty in the phase shifts at $s_0$ has been kept constant,
i.e. $3.4^\circ$ for the $S0$ wave. The central value of the phase changes
as $M_\pi$ decreases, partly because the shape of the imaginary part below
$s_A$ evolves, but primarily because the kinematic relation between phase
and imaginary part depends on the mass-dependent phase-space factor $\sigma(s)$.
As illustrated in the left panel of Fig.~\ref{fig:hyperp}, a fixed
uncertainty of $\pm 3.4^\circ$ in $\delta_0^0(s_0)$ translates into a
narrowing uncertainty band in $x_{It0}$. One might argue that keeping the
uncertainty in $x_{It0}$ constant would represent a more conservative 
choice. On the other hand, the unitarity limit
$|\sin \delta^0_0(s_A)| \leq 1$ implies the bound
$\Im t^0_0(s_A) \leq \sigma(s_A)^{-1}$, 
which is incompatible with an $M_\pi$-independent uncertainty in the
imaginary part, as also shown in Fig.~\ref{fig:hyperp}. In fact, the
unitarity bound even slightly impinges on the $\pm 3.4^\circ$
constant-phase uncertainty band. In view of this, maintaining a constant
uncertainty in the phase appears to be the most consistent choice.

\subsection{The $K\bar{K}$ threshold}
The assumption that the phenomenological input above $s_A$ does not change
significantly as we lower the light quark masses is adopted in this
analysis because we do not have any reliable way to calculate how the
$\pi\pi$ amplitude around $1\GeV$ or above depends on these masses. One
important feature, however, constitutes an exception:
the position of the $K\bar{K}$ threshold. How this depends on the light
quark masses can be predicted in \chpt~\cite{Gasser:1984gg} and the
relevant formulae have been worked out to NNLO (two
loops)~\cite{Amoros:1999dp}. At this threshold, the elasticity of the $S0$
wave drops sharply from unity (or very close to it), which is in large part
due to the presence of the $f_0(980)$ resonance, which couples strongly to
both $\pi \pi$ and $K\bar{K}$. This produces a sharp dip in the imaginary
part, visible in the left panel of Fig.~\ref{fig:imtI_central}. In view of
this, it would be unrealistic to leave the position of this dip fixed as  
the light quark masses decrease. On the other hand, while the quark-mass
dependence of the threshold position is well known, the exact deformation
of the shape of the imaginary part to the left and right of the dip cannot
be calculated model-independently. Here we adopt a pragmatic
approach and shift the position of the dip by using a kinematic map
of the variable $s$ in the region around the $K\bar{K}$ threshold. 

\begin{figure}[t]
  \centering
  \includegraphics[width=10cm]{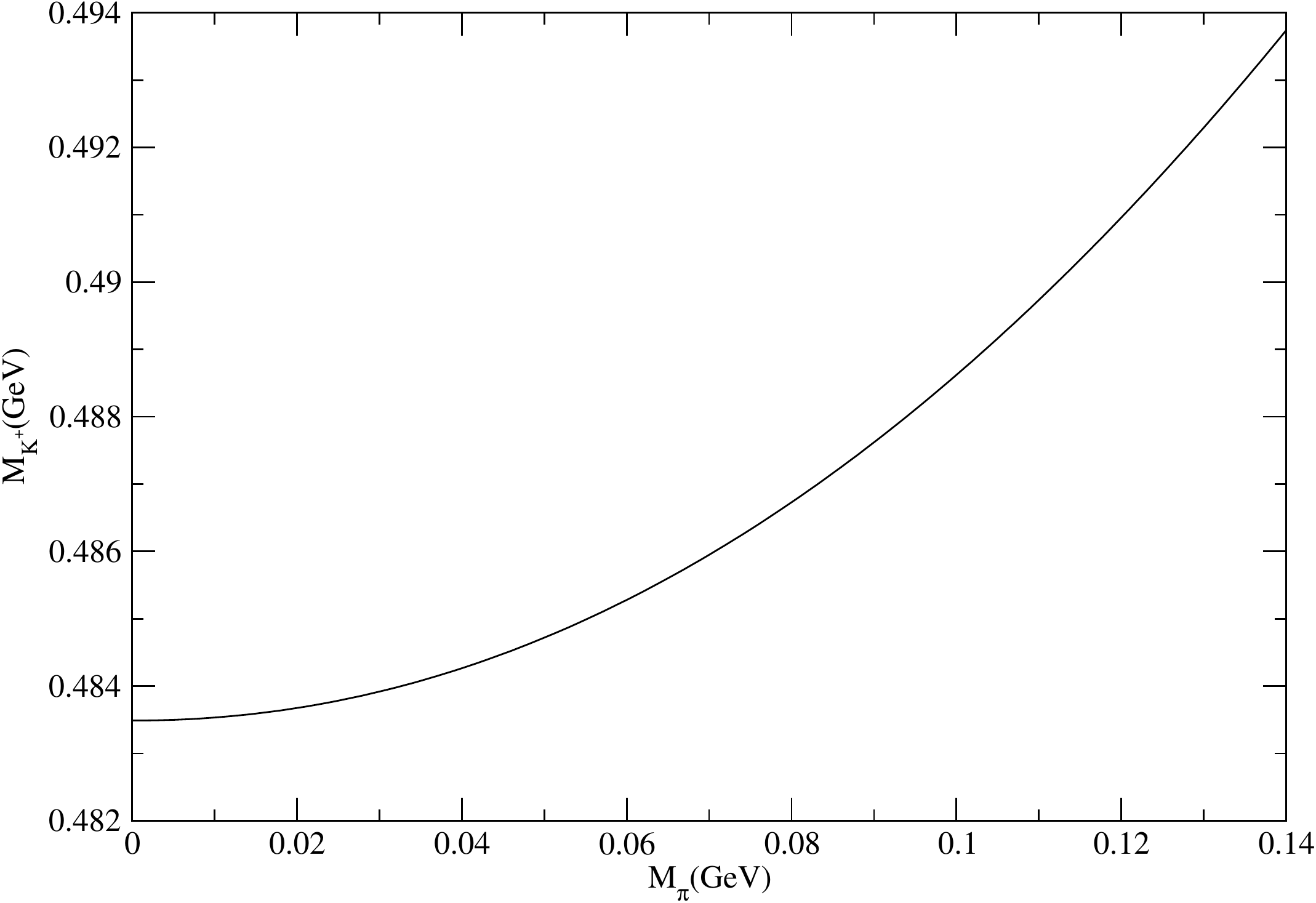}
  \caption{\label{fig:MKMpi} Dependence of $M_{K^+}$ on the light quark
    mass, expressed through the pion mass, in the region below its
    physical value, as predicted by \chpt~at NLO. }
\end{figure}
This mapping is defined as follows: for each value of $M_\pi$, the interval
$[s_A, 4 M_K^2(M_\pi^2)]$ is mapped to $[s_A, 4 M_{K,\text{phys}}^2]$, and
the interval $[4 M_K^2(M_\pi^2), s_2]$ is mapped to
$[4 M_{K,\text{phys}}^2,s_2]$, such that the boundary points $s_A$ and
$s_2$
remain fixed, and the first derivative of the mapping at both endpoints
equals unity. The explicit formula reads:
\begin{equation}
s_{\text{kin}}(s,M_\pi^2) = 
\begin{cases} 
s + \dfrac{4 (M_{K,\text{phys}}^2 - M_K^2)}{(4M_K^2 - s_A)^2} (s - s_A)^2 \, , & s_A \le s \le 4M_K^2, \\[14pt]
s + \dfrac{4 (M_{K,\text{phys}}^2 - M_K^2)}{(s_2 - 4M_K^2)^2} (s - s_2)^2 \, , & 4M_K^2 \le s \le s_2,
\end{cases}
\label{eq:smap_inv}
\end{equation}
with $M_K \equiv M_K(M_\pi)$. The input imaginary part $\Im t_0^0(s)$ for
all $s \geq s_A$ acquires an $M_\pi$ dependence through this mapping via
the replacement $\Im t_0^0(s) \to \Im
t_0^0\left(s_\text{kin}(s,M_\pi^2)\right)$.  For the numerical evaluation
we rely on one-loop \chpt, which is perfectly adequate for our
purpose. Using modern values of the LECs (see appendix~\ref{app:nums}) we
obtain the dependence shown in Fig.~\ref{fig:MKMpi}. As the pion mass
decreases from its physical value to the chiral limit, the charged kaon
mass shifts from $0.493677\GeV$ down to $0.4835\GeV$---a modest drop of
only about 2\%.

Accounting for this effect in this manner inevitably entails a degree of
model dependence: in particular, it assumes that the $f_0(980)$ dip and the
$K\bar{K}$ threshold remain locked together as we change the light-quark
masses, which can only be correct over a limited mass interval (for
a very recent analysis of the quark-mass dependence of the $f_0(980)$ and
other resonances, see Ref.~\cite{Pelaez:2026fmu}). On the other hand,
neglecting the threshold shift entirely would be unphysical. Given that the
shift is so small, any sensible parametrization  of this effect should
provide reliable results. We will assess the impact of this shift \textit{a
  posteriori} by comparing solutions obtained with or without it, adding 
the difference in quadrature to the uncertainty (with the only exception of
the shape of the imaginary part near the $K\bar{K}$ threshold, which is
treated differently, as discussed below). 

\subsection{Region of validity, inelasticities, and the
  case $M_\pi > M_{\pi,\mathrm{phys}}$}

The present discussion has so far ignored the fact that the limit of
validity of the Roy equations has been rigorously proven to be at least $68
M_\pi^2$. For $M_{\pi,\mathrm{phys}}$ this sits comfortably above $1\GeV$,
but as we move towards the chiral limit, the region of validity of the Roy
equations shrinks to zero. Since we solve the equations up to a fixed upper
boundary, $s_A = 0.84\GeV^2$, for all values of $M_\pi$, the limit of
validity falls below $s_A$ as soon as $M_\pi < 111\MeV$. Is this a problem?
The boundaries of the region of validity are linked to the convergence of
the partial wave expansion: all angular integrals needed for the projection
onto partial waves must avoid the double-spectral region, otherwise the
resummation of partial-wave series will not converge. In the derivation of
the Roy equations the partial wave expansion is applied twice: once
externally, on the amplitude represented as a fixed-$t$ dispersion relation
and once internally, for the imaginary part of the amplitude under the
integral sign. The first, external expansion is unproblematic for any
energy, if one does not try to resum the series of partial waves (which we
are not doing here). The second, internal expansion is more critical,
because what appears in the integrand is the total amplitude: if one
replaces it by its partial-wave expansion, the series must converge. In
particular, deriving the equations requires interchanging the partial-wave
sum with the angular integral of the external projection, an operation that
is rigorously justified only for a convergent series.

In our practical treatment of the Roy equations, however, this formal issue
does not pose a concrete problem. In the region between threshold and $s_2$
we do expand the total amplitude under the integral sign, but only retain
the first few terms in each channel, up to $D$ and $F$ waves, relying on
the phenomenological fact that these few waves approximate the full
amplitude well. Above $s_2$ we do not rely on the partial wave expansion
and use a Regge representation for the total amplitude (for details, see
Ref.~\cite{Caprini:2011ky}). Provided that the phenomenology in the region
between $1$ and $2\GeV$ does not change significantly as the pion mass
decreases, this pragmatic approach will not break down as we approach the
chiral limit.\footnote{For an indirect confirmation of this, see the
discussion and the figures in Sect.~10 of
Ref.~\cite{Ananthanarayan:2000ht}, which concern the behaviour of the Roy
integrals above $s_0$. They show that there is no visible disruption as one
crosses the $1.15\GeV$ boundary.}

A complementary line of argument leads to the same conclusion.  From the
chiral point of view, double-spectral regions arise at the three-loop
level, or $\cO(p^8)$. The same is true for $4 \pi$ inelasticities, whose
threshold also moves towards zero as we approach the chiral limit. Since
the chiral input we are using includes contributions up to $\cO(p^6)$ it is
justified to neglect these higher-order effects within the domain of
validity of \chpt. Outside this region we can rely on phenomenology: this
shows very clearly that multipion inelastic effects are essentially
negligible unless they are enhanced by resonances. Indeed, although the
$4\pi$ inelastic threshold is at $0.56\GeV$, the corresponding inelasticity
has been neglected in all previous Roy-equation analyses, even up to
$1.15\GeV$. We therefore assume that neither effect compromises the
validity of the Roy equations in the present implementation even as we
decrease the pion mass all the way to zero. As shown below, the solutions
obtained in this limit can be tested, providing an indirect check of this
and other assumptions made so far.

We stress that when moving in the opposite direction, the situation changes
completely: increasing the light-quark masses pushes both the inelastic
threshold and the limit of validity of the Roy equations to higher
energies. While this appears to be an advantage, there are other causes for
concern: first of all, the convergence of the chiral expansion for the
scattering lengths is known to become problematic once the pion mass goes
above $200\MeV$~\cite{Colangelo:2001df} (at least for $a_0^0$). Matching
with \chpt\ below threshold, as done in~\cite{Colangelo:2001df}, improves
the convergence, but only up to a certain point. Second, the input above
$1\GeV$ will start, at some point, to change significantly as the pion mass
increases, which would require a dedicated study.

\section{Numerical solutions}
\label{sec:numsol}
\begin{figure}[t]
  \centering
  \includegraphics[width=7.5cm]{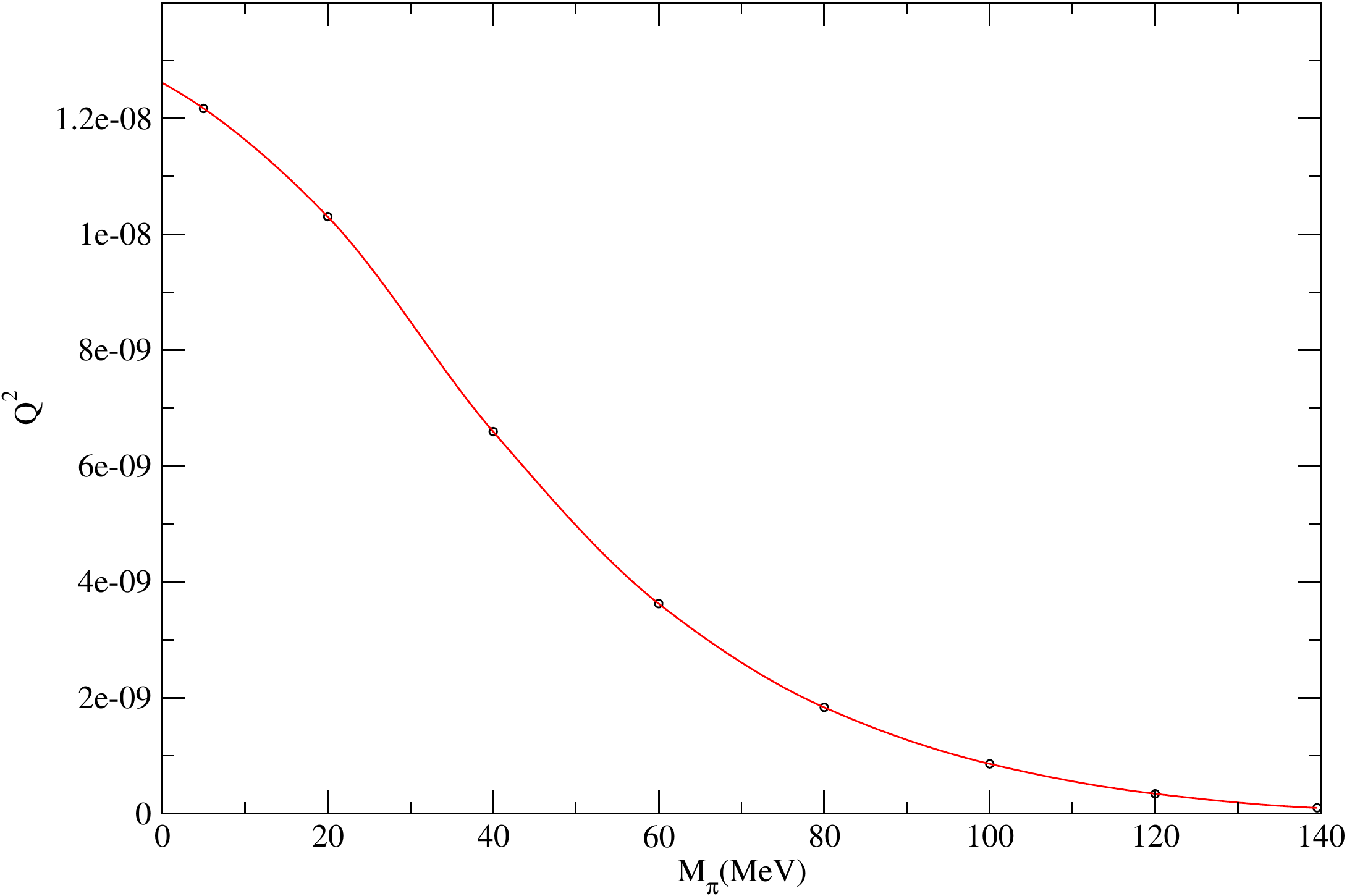}
  \includegraphics[width=7.5cm]{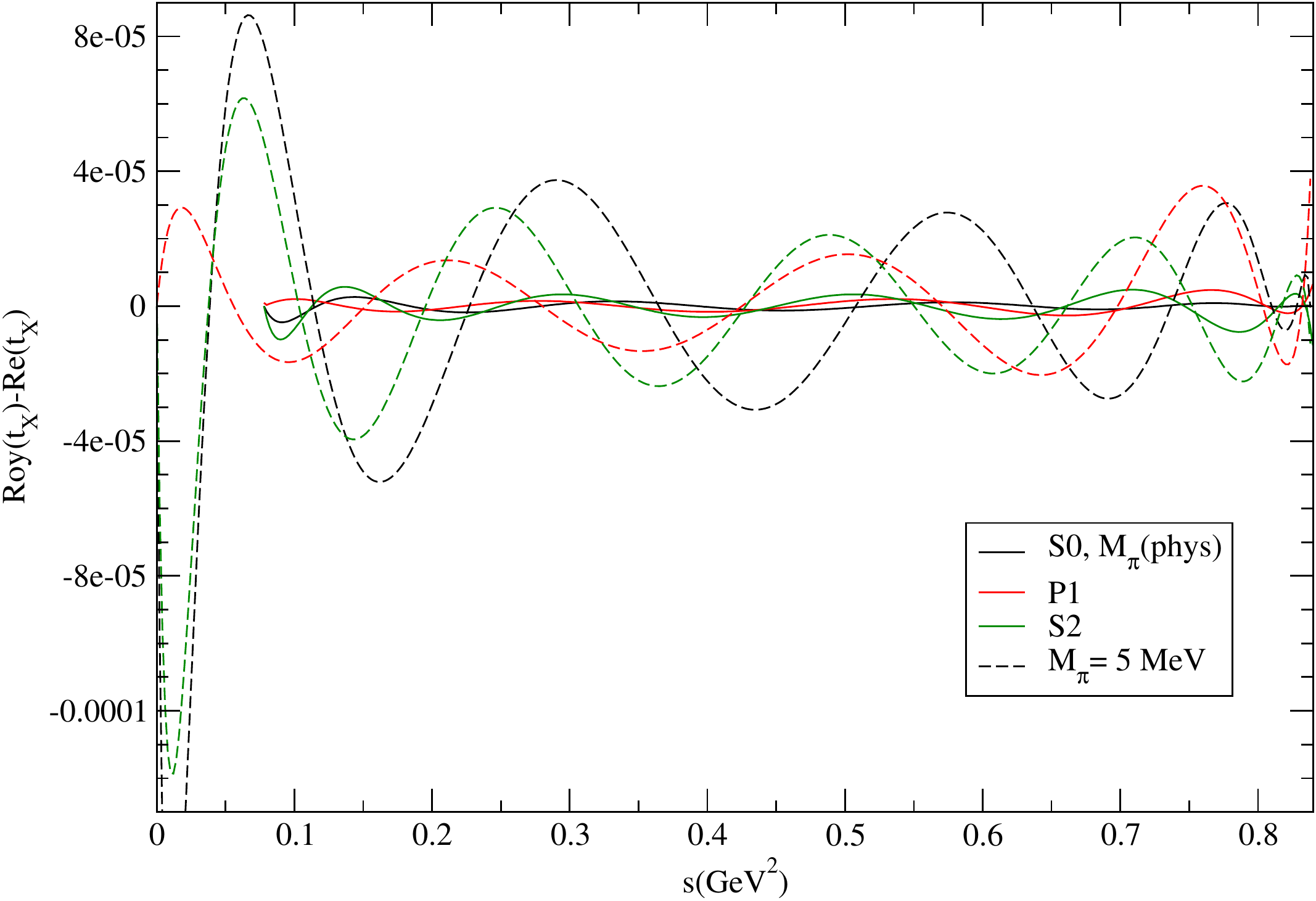}
  \caption{\label{fig:SoS} Left: $Q^2$ values for the central solutions
    ($x_{It0}=0$) as a function of the pion mass. The curve is a polynomial
    interpolation through the eight $Q^2$ values of the calculated
    solutions, which are shown as small circles. Right: difference between
    the right- and left-hand sides of the Roy equations for the physical
    and the $5\MeV$ pion mass. The negative peak of the $S0$ wave for the
    $5\MeV$ case (not shown here in order not to compress the vertical
    scale of the plot) reaches $-2.1\cdot 10^{-4}$. }
\end{figure}
Within this framework, solutions of the Roy equations at any value of
$M_\pi < M_{\pi,\mathrm{phys}}$ can be found, and they are of excellent to
very good quality. In the following, solutions at seven different values
below the physical pion mass are obtained: from 120 to $20\MeV$ in $20\MeV$
intervals, and a final one at $5\MeV$. As anticipated above, the $Q^2$,
which provides a measure of the quality of the solution, degrades by about
two orders of magnitude between the physical and the lowest pion mass, as
shown in the left panel of Fig.~\ref{fig:SoS}, but the highest $Q^2$ is
still indicating a solution of very high quality. This is more explicitly
seen in the right panel of Fig.~\ref{fig:SoS} which shows the difference
between the right-hand and the left-hand side of the Roy equations for the
three waves and for the physical and $5\MeV$ masses. In the $5\MeV$ case
there is a negative peak of $\cO(10^{-4})$ in both $S$ waves, but otherwise
the typical deviation is a few $10^{-5}$. For the physical mass solution
the deviation is $\cO(10^{-5})$, but mostly even smaller.

\begin{figure}[t]
  \centering
  \includegraphics[width=14cm]{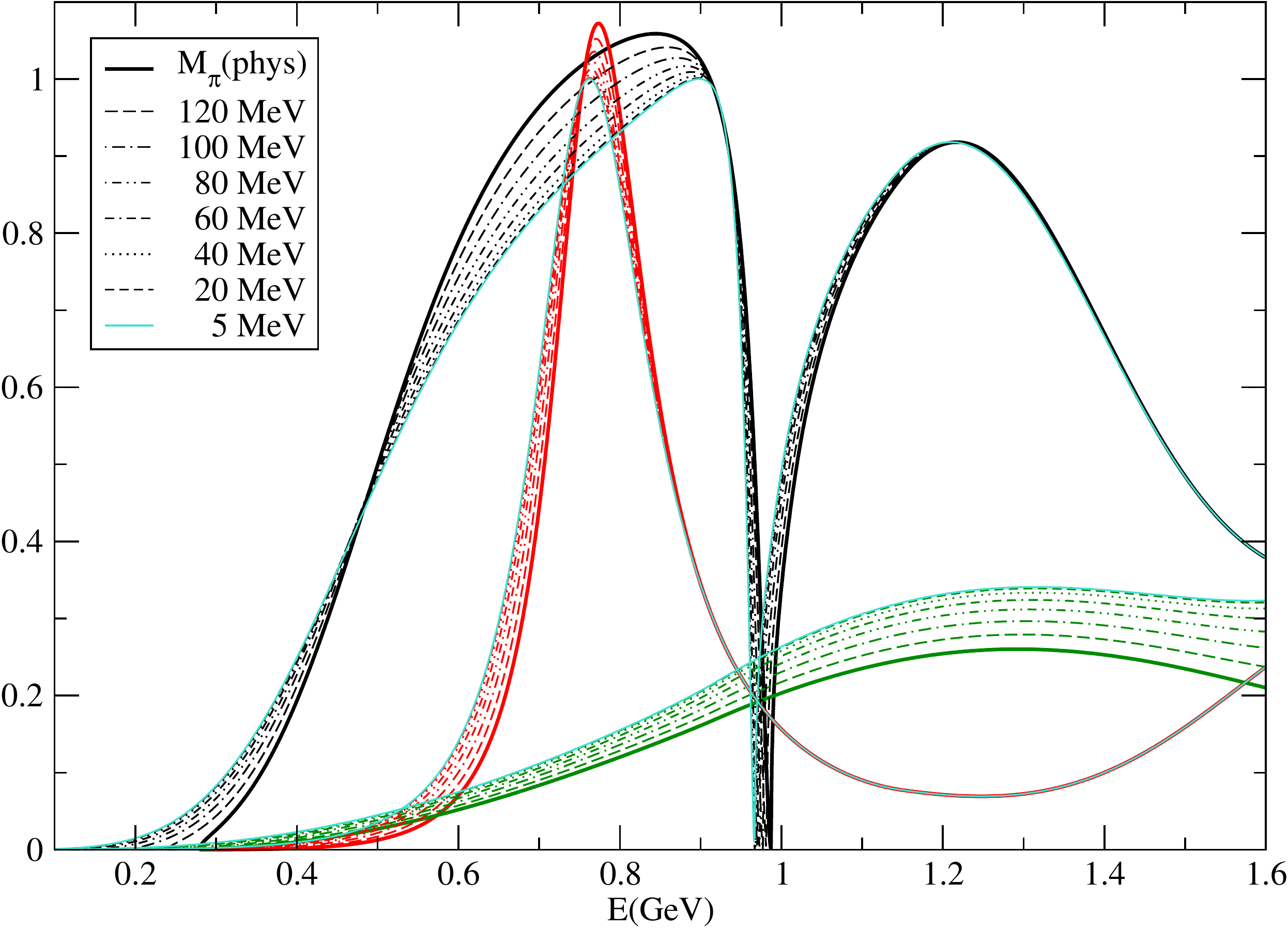}
  \caption{\label{fig:ImtI_Mpi} Imaginary parts of the Roy solutions for
    the lowest partial waves (black: $S0$, red: $P$, green: $S2$), for eight
    values of the pion mass.}
\end{figure}

The imaginary parts of the Roy solutions for the three lowest waves and the
eight values of the pion mass considered are shown in
Fig.~\ref{fig:ImtI_Mpi}. 
The figure shows very clearly what the main change at low energy is: the
$2\pi$-threshold shifts towards zero and the scattering lengths decrease,
making the start of the imaginary part flatter as we move towards 
the chiral limit. The figure also makes visible the features of our
framework in the input region (above $s_A$) as defined above:
the $P$ wave is kept strictly constant with the mass, whereas for the $S0$
wave the only change is the shift of the $K\bar{K}$-threshold as predicted
by NLO \chpt. For the $S2$ wave we have allowed for a rigid shift of the
whole curve between $s_A$ and $s_2$, following the approach adopted
in~\cite{Ananthanarayan:2000ht,Caprini:2011ky}, even though there is no
reason for the behaviour of the $S2$ wave above $1\GeV$ to depend on the light
quark masses: the dependence of $\Im t^2_0(s)$ on the pion mass above $1\GeV$
shown in the figure is not meant to be taken literally and only illustrates
the present setup. As Fig.~\ref{fig:imtI_central} shows, the uncertainties
of the data in that region are large and an explicit calculation has shown
that changing the behaviour of $\Im t^2_0(s)$ between $s_A$ and $s_2$ has
only a minute impact on the solutions below $s_A$.\footnote{A shift in $\Im
t^2_0(s)$ above $s_A$, in the form of a quadratic polynomial which vanishes
together with its first derivative at $s_A$ while reaching 0.1 at $s_2$,
generates a maximal change below $s_A$ for each of the eight pion masses of
about ($2\cdot 10^{-4}$, $6 \cdot 10^{-4}$, $2\cdot 10^{-3}$) for ($\Im
t^0_0$, $\Im t^1_1$, $\Im t^2_0$). Adding the shifted curves to 
Fig.~\ref{fig:ImtI_Mpi} would make them barely visible for $\Im t^2_0$ and
invisible for the other two.}

\begin{figure}[t]
  \centering
  \includegraphics[width=7.4cm]{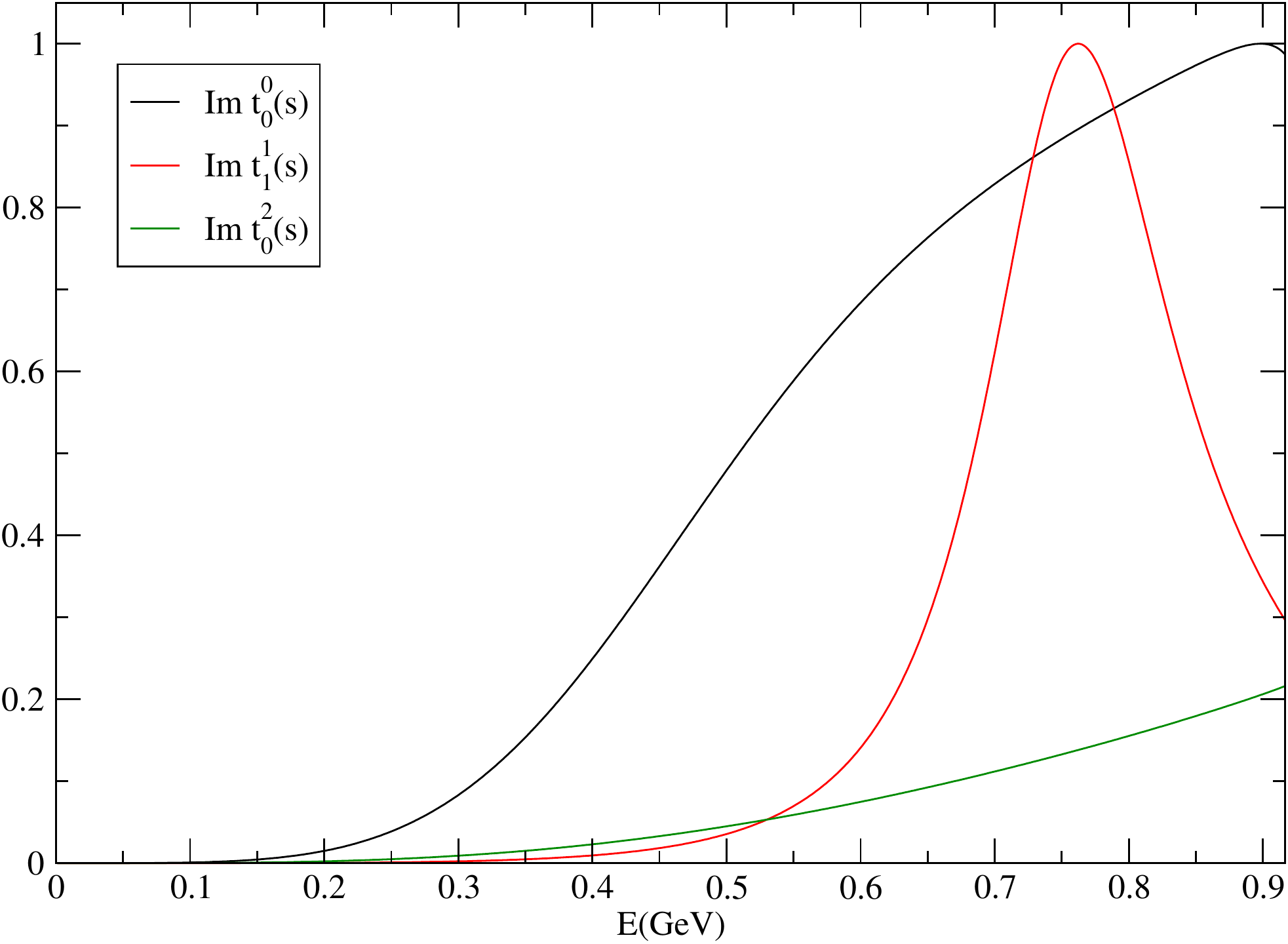}
  \includegraphics[width=7.4cm]{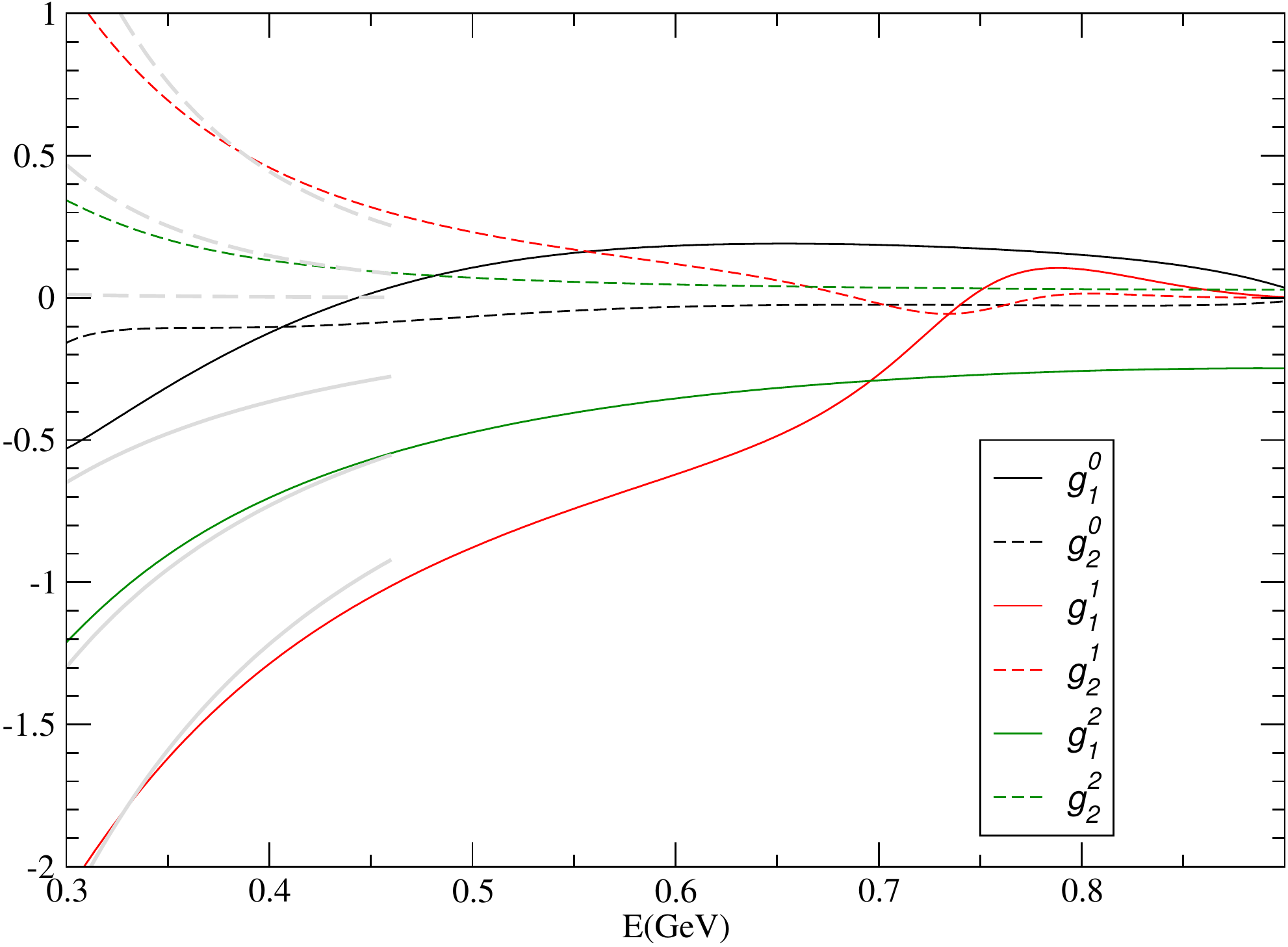}
  \caption{\label{fig:Mpicoeffs} Left: Imaginary parts of the three lowest
    partial waves below $s_A$ in the chiral limit. All three waves would be
    indistinguishable from the $5\MeV$ solution shown in
    Fig.~\ref{fig:ImtI_Mpi}. Right: Coefficients describing the pion mass
    dependence of the three lowest imaginary parts, in the region between
    $0.3\GeV$ and $s_A$. The light-grey curves below $0.46\GeV$ represent
    the chiral predictions, Eq.~\eqref{eq:gIchpt}.}  
\end{figure}

The figure also shows that the changes of the imaginary parts with the pion
mass are very smooth and in first approximation can be well described by a
linear polynomial in $M_\pi^2$, so linear in the light quark masses. Moving
up to a quadratic polynomial in $M_\pi^2$ one obtains an almost perfect
description of the pion mass dependence of the imaginary parts across the
whole energy region between threshold and $s_A$, so well beyond the chiral
regime. To be more precise, fitting the $M_\pi$ dependence of
the imaginary parts with
\begin{equation}
\Im t^I_\ell(s,M_\pi)=\Im t^I_\ell(s,0)\left[1+g^I_1(s) y_\pi + g^I_2(s)
  y_\pi^2 \right] \, ,
\label{eq:yexp}
\end{equation}
where $y_\pi:=(M_\pi/M_{\pi,\text{phys}})^2$, yields maximal relative
residuals below $1\%$ for all values of $\sqrt{s}$ above
$0.3\GeV$.\footnote{With the single exception of the $P$ wave, which gives
a maximal relative residual of about $3\%$ below $0.4\GeV$. The wave itself
is already tiny at that energy, so that the absolute deviation remains
negligible.}  A plot of the imaginary parts in the chiral limit and of the
six coefficients is shown in Fig.~\ref{fig:Mpicoeffs}. The imaginary parts
in the chiral limit are almost indistinguishable from the $5\MeV$ solution
shown in Fig.~\ref{fig:ImtI_Mpi}, with maximal differences between the
chiral and the $5\MeV$ solution of $2\cdot 10^{-4}$ for the $S0$ and $P$
waves and $3\cdot 10^{-5}$ for the $S2$ wave. As for the $g^I_i(s)$
coefficients, the plot shows that they are rather well-behaved functions,
with some growth towards threshold, which we explain below, and a
mild oscillation in the $P$ wave in the $\rho$ region. That such a
good description of the $M_\pi$ dependence is possible even well beyond the
chiral regime is, of course, not a surprise, and mainly related to the fact
that the light quark masses are very small compared to the QCD scale. Any
QCD observable can be expanded in these masses and only a few terms are
expected to be needed to describe the quark mass dependence between the
chiral limit and the physical value. The fact that the $g^I_2$ coefficients
are smaller than the $g^I_1$ also indicates that this expansion is probably
going to work well even beyond the physical mass. This represents a
possible path to reach the region of pion masses above the physical value,
which is the one of interest for lattice applications. A more direct and
promising approach consists in solving the Roy equations for $M_\pi >
M_{\pi,\mathrm{phys}}$ as will be done in Sect.~\ref{sec:heavierMpi}.

Even if this extrapolation to heavier pion masses will not be used here, it
is nevertheless worthwhile making the discussion more precise, clarifying
an apparent puzzle. Indeed one could argue that for the expansion in
$M_\pi$ to be well-behaved not only the ratio $g^I_2/g^I_1$ needs to be
small, but also the coefficient $g^I_1$ itself. Figure~\ref{fig:Mpicoeffs}
shows that all six coefficients grow as inverse powers of $s$ in the
low-energy region, three of them becoming larger than unity in magnitude
already above $0.3\GeV$. The point is that in the region below $\sim
0.5\GeV$ we can rely on the chiral counting, which states that $\Im
t^I_\ell(s)= \cO(p^4)$ and that all three terms in square brackets in
Eq.~\eqref{eq:yexp} are of the same chiral order. Their explicit
expressions are easily obtained and read
\begin{align}
  \label{eq:gIchpt}
  g^0_1(s)&=-\frac{3}{\hat{s}} \, , \quad g^1_1(s)=-\frac{10}{\hat{s}} \, ,
  \quad g^2_1(s)=-\frac{6}{\hat{s}} \,, \nonumber \\
  g^0_2(s)&=\frac{1}{4\hat{s}^2} \, , \quad
 \quad g^1_2(s)=\frac{30}{\hat{s}^2} \, , \quad g^2_2(s)=\frac{10}{\hat{s}^2} \, ,
\end{align}
where $\hat{s}:=s/M_{\pi,\mathrm{phys}}^2$. This is the result of an exact
expansion around $M_\pi=0$, whereas our numerical determination of the
$g^I_n$ coefficients is based on a fit of the solutions with nonzero
masses, which is bound to be inaccurate in the threshold region.
Nonetheless, as shown in Fig.~\ref{fig:Mpicoeffs}, right panel, the fitted
coefficients $g^I_n$ follow the LO chiral behaviour given in
Eq.~\eqref{eq:gIchpt} quite closely in the low-energy region. For the $S0$
wave the agreement is not better than qualitative, reflecting the fact that
in this wave threshold effects are more prominent than in the other waves.
This shows that, in this region, we do
not gain more insights into this issue if we rely on our numerical
expansion rather than on a pure \chpt~analysis, and this has already been
provided in the past~\cite{Bijnens:1995yn,
  Bijnens:1997vq,Colangelo:2001df}. In particular,
in~\cite{Colangelo:2001df} (see Fig.~11) it has been shown that the
convergence rate differs among partial waves, and that for the $S0$ wave at
threshold the convergence becomes very slow already for a pion mass of
$200\MeV$.

Above $0.5\GeV$ we enter a different regime, where \chpt\ no longer provides 
guidance. In this region we can perform a Taylor expansion in
$M_\pi^2$, evaluate the coefficients numerically with our Roy solutions,
and assess the convergence on the basis of their ratios. It is in this
region that the expansion in $M_\pi^2$ works well for any pion mass below
the physical value only if both $|g^I_1| \ll 1$ and $|g^I_2/g^I_1| \ll
1$. Figure~\ref{fig:Mpicoeffs} shows that this is indeed the case if one
stays away from the zeros of the $g^I_1$ coefficients. On the basis of these
coefficients one can also assess if the expansion in $M_\pi^2$ remains
applicable even above the physical pion mass, and up to which mass value. This
can be done by calculating the value of the pion mass for which the ratios
$|g^I_1 y_\pi|$ and $|g^I_2 y_\pi/g^I_1|$ become close to one. A conservative
estimate of this upper limit is that it should be around $300\MeV$. It is
interesting that this upper limit is higher than the one obtained for
the threshold region on the basis of \chpt.

\subsection{Uncertainties}
\label{sec:Unc}
The uncertainty estimate in Ref.~\cite{Caprini:2011ky} was a complex
endeavour. The low- and the high-energy regions of the $\pi \pi$ scattering
amplitude are linked by several constraints: continuity in the transition
region for each isospin amplitude and energy-dependent sum rules involving
integrals from threshold up to infinity. After varying each independent
input parameter within uncertainties, the whole system of constraints was
satisfied again and the Roy equations for all waves, up to $D$ and $F$, were
solved anew in an iterative procedure, until convergence was reached. Only
at this point was the variation of a given observable (like the imaginary part
of a given wave in the region where the Roy equations are solved) with
respect to that input parameter evaluated. Adding these up 
in quadrature provided the total uncertainty.

\begin{figure}[t]
  \centering
  \includegraphics[width=15cm]{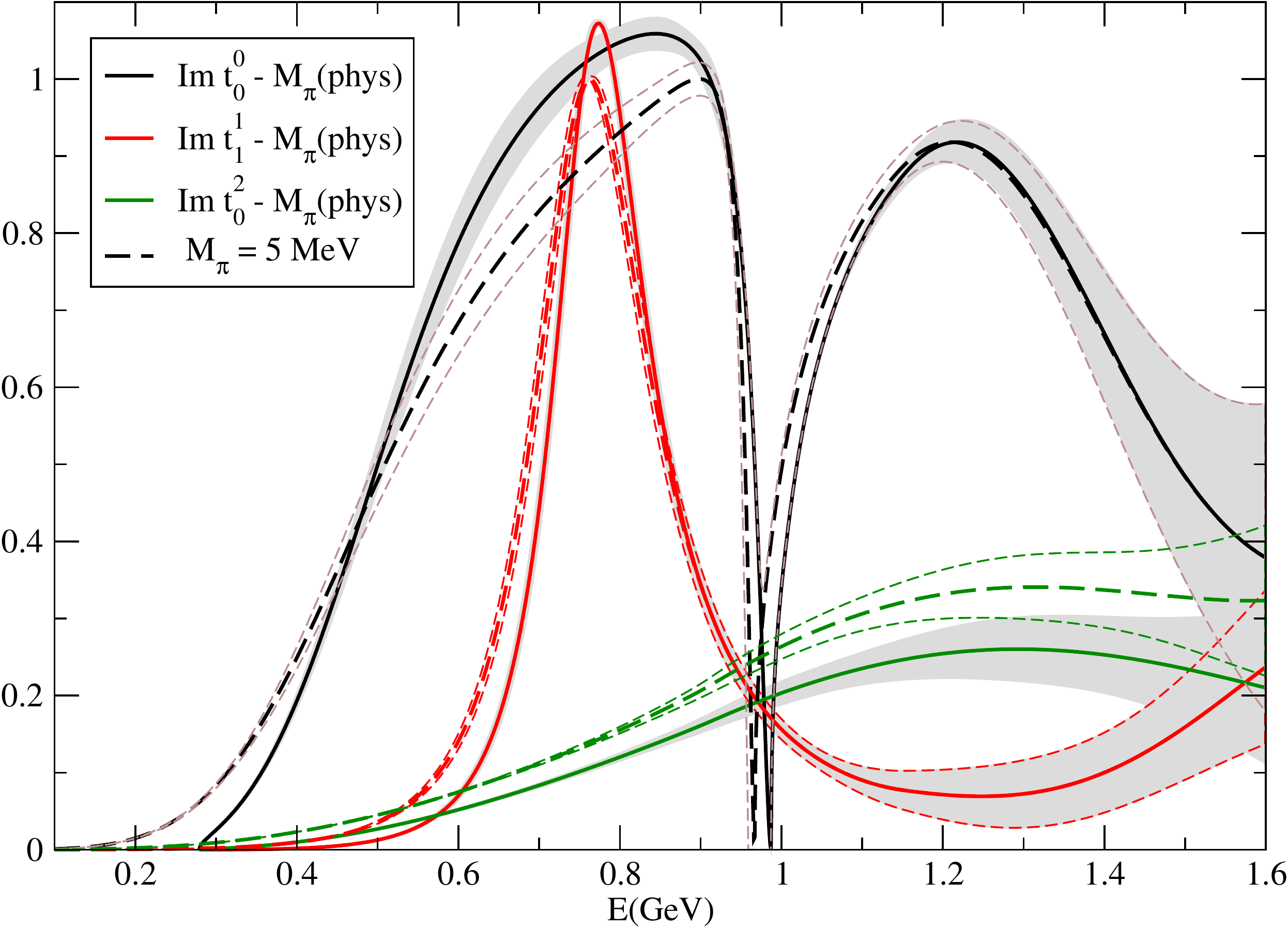}
  \caption{\label{fig:ImtI_unc} Uncertainty bands for the imaginary parts
    of the three lowest partial waves for the physical and $5\MeV$ pion
    masses. As discussed above, the $5\MeV$ curves are essentially
    indistinguishable from the extrapolation to the chiral limit, so the
    uncertainty bands shown here well represent those in the chiral limit.}  
\end{figure}
In the present work, a simplified procedure is adopted. In addition to the
uncertainty in $\Im t_0^0(s_A)$ which we have discussed above, we treat the
uncertainty in $\Im t_1^1(s_A)$ similarly, linking it to the uncertainty in
$\delta_1^1(s_0)$. This had been estimated to be equal to $\pm2^\circ$ in
Ref.~\cite{Ananthanarayan:2000ht}, a value we adopt here (even
though more recent analyses of the $e^+e^- \to \pi^+ \pi^-$
data~\cite{Colangelo:2018mtw} show that these are precise enough to
constrain this phase even further). We then have the uncertainties for the
two $S$-wave scattering lengths: $a_0^0=0.220 \pm 0.005$ and $a_0^2=-0.0444
\pm 0.0010$, which we scale to lower $M_\pi$ values with the scaling
factor $y_\pi^2$.\footnote{This means that we overestimate them towards the
chiral limit, since part of them is due to $\cO(p^6)$ contributions which
go even faster to zero.}

\begin{figure}[t]
  \centering
  \includegraphics[width=15cm]{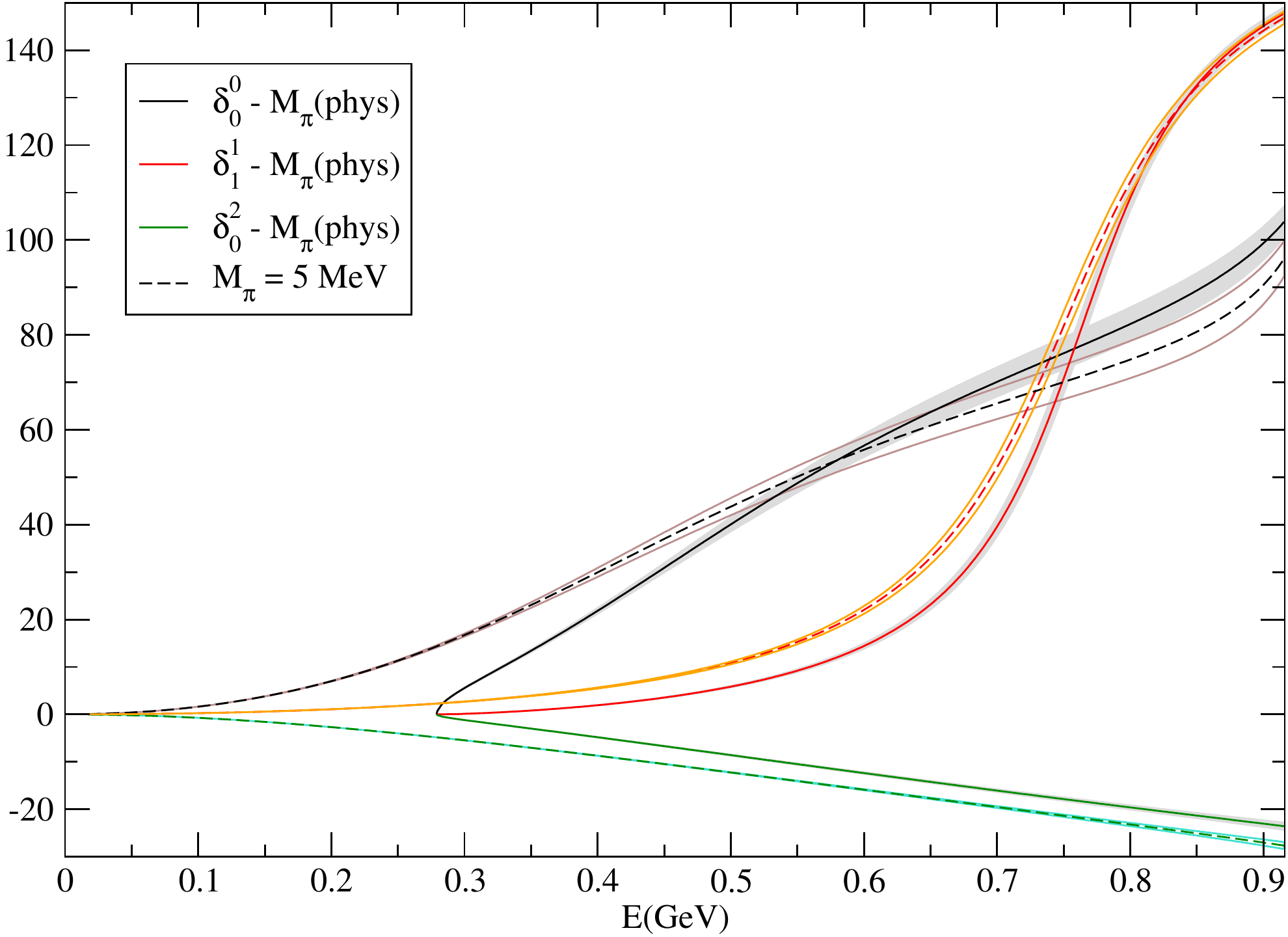}
  \caption{\label{fig:phases_unc} Uncertainty bands for the phase shifts 
    of the three lowest partial waves for the physical and $5\MeV$ pion
    masses. As discussed above, the $5\MeV$ curves are essentially
    indistinguishable from the extrapolation to the chiral limit, so the
    uncertainty bands shown here well represent those in the chiral limit.}  
\end{figure}
As for the region above $s_A$, some simplifications are also adopted. For
each of the lowest partial waves we vary the value at $s_2$ between
generous ranges: $\pm 0.2$ for the $S0$ wave and $\pm 0.1$ for the other
two, as these roughly reflect the current experimental situation. These
variations are implemented with the help of a second-degree polynomial
which multiplies the given uncertainty and has the properties that it
vanishes, together with its first derivative at $s_A$, and becomes equal to
unity at $s_2$. In the region between $s_A$ and $s_2$, we also
consider an independent source of uncertainty, treating the $S0$ wave
differently from the other two. For the $S0$ wave, we shift the position of
the $f_0(980)$ resonance and the $K\bar{K}$ threshold according to the \chpt\
prediction, taking the difference with respect to the unshifted threshold
as a measure of systematic uncertainty. Because this threshold shift already
affects the surrounding region to both the left and right, we do not add any
further source of uncertainty around $1.2\GeV$ in this channel. For the
other two waves, however, we allow the imaginary part at $1.2\GeV$ to vary
independently of the boundary values at $s_A$ and $s_2$, by an additional
$\pm 0.03$ and $\pm 0.034$ for the $P$ and $S2$ waves,
respectively\footnote{These values have been chosen in such a way that the
whole uncertainty band grows smoothly and without bumps towards higher
energies.}. We then consider the uncertainties in the driving terms, but
instead of dealing with each individual source, we take them \textit{en bloc}
and vary all of them simultaneously by one standard deviation.

For each observable, the individual uncertainties are added in quadrature,
with the sole exception of the systematic uncertainty associated with the
$K\bar{K}$ threshold shift. For example, the resulting uncertainty bands
for the imaginary parts of the three lowest waves for the physical and for
the $5\MeV$ mass (which essentially coincides with the chiral limit) are
shown in Fig.~\ref{fig:ImtI_unc}. Concerning the shift of 
the $K\bar{K}$ threshold, adding in quadrature the difference between the
central curve and the one without any shift would generate unphysically
sharp spikes in the uncertainty. Instead, we construct an envelope of the two curves. In
particular, we allow the zero of the imaginary part to lie anywhere in the
interval between $0.96$ and $0.98\GeV$, as can be seen clearly in
Fig.~\ref{fig:ImtI_unc}. An analogous plot for the phase shifts is shown in
Fig.~\ref{fig:phases_unc}. In this case the uncertainty due to the handling
of the $K\bar{K}$ threshold does not produce any localized spikes and
requires no special treatment.

\section{Threshold parameters: a test of the chiral behaviour of the solutions}
\label{sec:disp-chpt}
The threshold parameters of the $S$, $P$ and higher waves can be calculated
via sum rules with the Roy-equation solutions as
input, as first shown by Wanders~\cite{Wanders:1966}. To obtain these sum
rules one simply needs to expand the Roy representation of the partial
waves near threshold. The explicit formulae for these sum rules have been
provided in modern notation in Ref.~\cite{Ananthanarayan:2000ht} and are
reproduced here for convenience, after some simplifications, first for the
$S$ and $P$ waves:
\begin{align}
  \label{eq:sumruleSP} 
 b_0^0  &=  \frac{1}{3 M_\pi^2} (2 a_0^0-5 a_0^2) 
+ \frac{16 M_\pi^2 }{3\pi}\int_{4 M_\pi^2}^{s_2} \frac{ds}{s^2(s-4
  M_\pi^2)}\left\{ 4 \Im t_0^0(s)- 9 \Im t_1^1(s)+ 5 \Im t_0^2(s) \right\}
\nonumber \\
&+\frac{64 M_\pi^4}{\pi}\int_{4 M_\pi^2}^{s_2}\!\!\!\! ds\frac{[\Im
    t_0^0(s) - \sigma(s) (a_0^0)^2]}{[s(s-4M_\pi^2)]^2} 
-\frac{16(a^0_0)^2(s_2^2-2 s_2 M_\pi^2-2M_\pi^4)}{3 \pi s_2^2 M_\pi^2 \sigma(s_2)}+b^0_{0\, d} \; , \nonumber \\ 
b^2_0&=-\frac{1}{6 M_\pi^2} (2a_0^0-5a_0^2)  + \frac{8 M_\pi^2 }{3\pi}\int_{4
  M_\pi^2}^{s_2} \frac{ds}{s^2(s-4M_\pi^2)}\left\{ 2  \Im t_0^0(s) +9 \Im
t_1^1(s) + 7 \Im t_0^2(s) \right\} \nonumber \\
& +\frac{64 M_\pi^4}{\pi}\int_{4 M_\pi^2}^{s_2} \!\!\!\! ds \frac{[\Im
    t^2_0(s)-\sigma(s) (a_0^2)^2]}{[s(s-4M_\pi^2)]^2} 
-\frac{16(a^2_0)^2(s_2^2-2 s_2 M_\pi^2-2M_\pi^4)}{3 \pi s_2^2 M_\pi^2 \sigma(s_2)} +b^2_{0\,d} \; , \nonumber \\  
a^1_1&= \frac{1}{18 M_\pi^2}\,(2a^0_0-5a^2_0)
-\frac{8 M_\pi^2}{9\pi}\int_{4M_\pi^2}^{s_2} \frac{ds}{s^2(s-4 
  M_\pi^2)}\left\{2 \Im t_0^0(s) - 9 \Im t_1^1(s) - 5 \Im
t_0^2(s)\right\} \nonumber \\
& + \frac{16 M_\pi^2}{\pi}\int_{4M_\pi^2}^{s_2}ds \frac{\Im t_1^1(s)}{s(s-4 
  M_\pi^2)^2} +a^1_{1\,d} \; , \\  
b^1_1&= \frac{8 }{9\pi}\int_{4 M_\pi^2}^{s_2}
\frac{ds}{s^3}\left\{-2 \Im t_0^0(s) +9 \Im t_1^1(s) + 5 \Im
t_0^2(s)\right\} + \frac{16 }{\pi}\int_{4 M_\pi^2}^{s_2} \!\!\!\! ds \frac{\Im t_1^1(s)}{(s-4
  M_\pi^2)^3} + b^1_{1\,d}  \;, \nonumber 
\end{align}
and then for the two $D$-wave scattering lengths:
\begin{align}
  \label{eq:sumruleD}
  a^0_2&= \frac{16}{45\pi} \int_{4M_\pi^2}^{s_2} 
\frac{ds}{s^3 } \left\{ \Im t_0^0(s)+9 \Im\,t_1^1(s)+ 5\Im t_0^2(s)
\right\} + \frac{128 M_\pi^2}{5\pi} \int_{4M_\pi^2}^{s_2} \!\!\!\! ds 
\frac{\Im t_1^1(s)}{s^3 (s-4M_\pi^2)} +a^0_{2\,d} \; , \nonumber \\
a^2_2 &=\frac{8}{45\pi} \int_{4M_\pi^2}^{s_2} 
\frac{ds}{s^3 } \left\{2 \Im t_0^0(s)- 9 \Im\,t_1^1(s)+ \Im t_0^2(s)
\right\} -\frac{64 M_\pi^2}{5\pi} \int_{4M_\pi^2}^{s_2} \!\!\!\! ds 
\frac{\Im t_1^1(s)}{s^3 (s-4M_\pi^2)} +a^2_{2\,d} \; ,
\end{align}
where $a^I_{\ell,d}$ and $b^I_{\ell,d}$ represent the driving-term
contributions to the sum rules. It is instructive to apply the chiral counting
to these expressions: the $S$-wave scattering lengths start at $\cO(p^2)$
and are the only contribution of this chiral order, whereas the imaginary
parts (and therefore the integrals) start at $\cO(p^4)$. This implies that
of the six threshold parameters shown here, the first (last) three start at
$\cO(p^2)$ ($\cO(p^4)$). This is relevant for the comparison we are going
to make.

In Ref.~\cite{Colangelo:2001df} these expressions have been used to
calculate the threshold parameters after the matching between the chiral
and the dispersive representation, arguing that this way of calculating
them is more reliable than the direct calculation in \chpt. A clear
indication of this is provided by $a_0^0$, which shows a much better
convergence behaviour if one calculates it with the dispersive
representation after a matching below threshold, rather than calculating it
directly at threshold in two-loop \chpt. Nonetheless, the numbers provided
in Table~2 of Ref.~\cite{Colangelo:2001df} show that the two ways of
calculating the threshold parameters agree well, mostly within one sigma,
with just a few exceptions.

\begin{figure}[t]
  \centering
  \includegraphics[width=8.9cm]{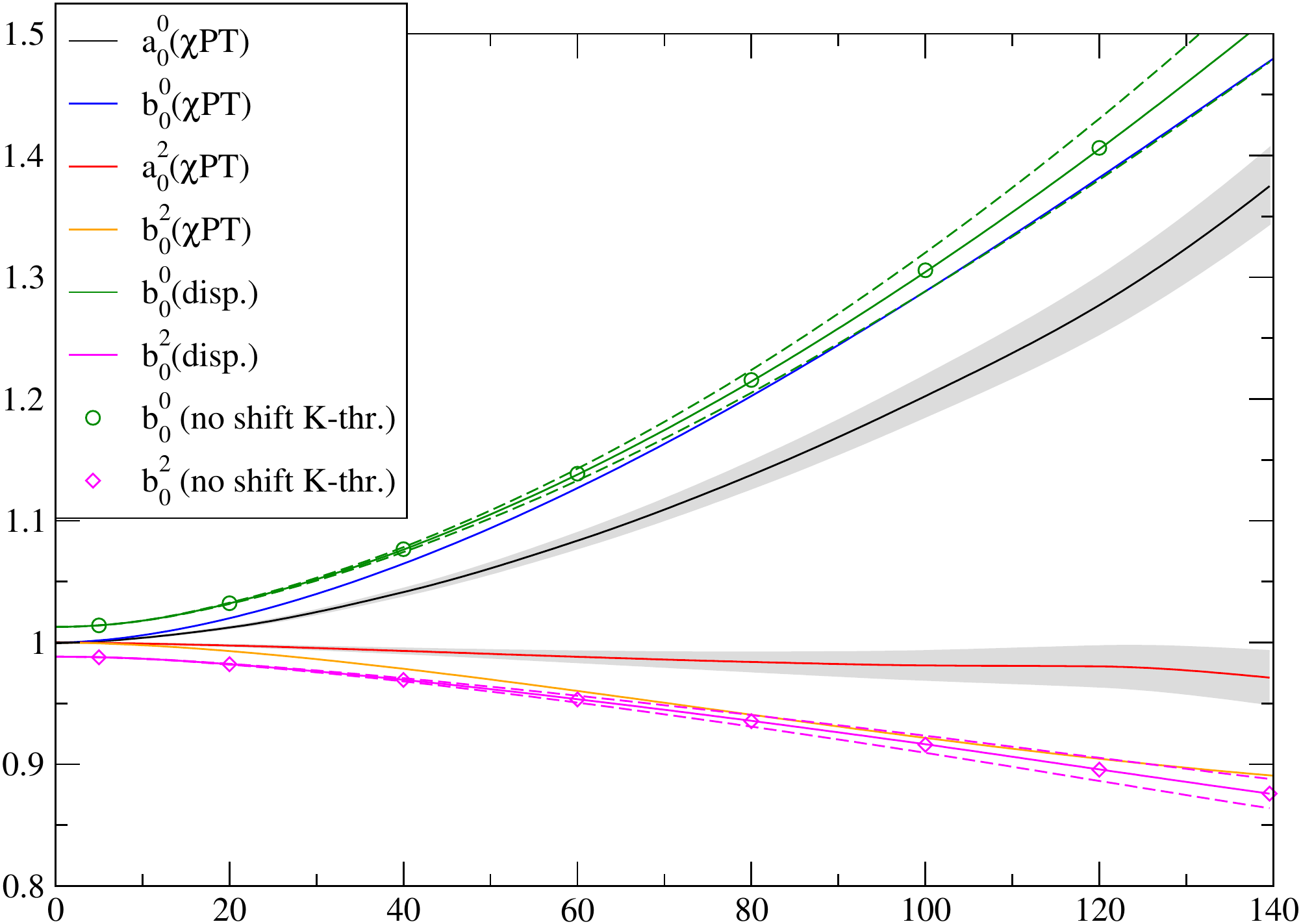}
  \includegraphics[width=8.9cm]{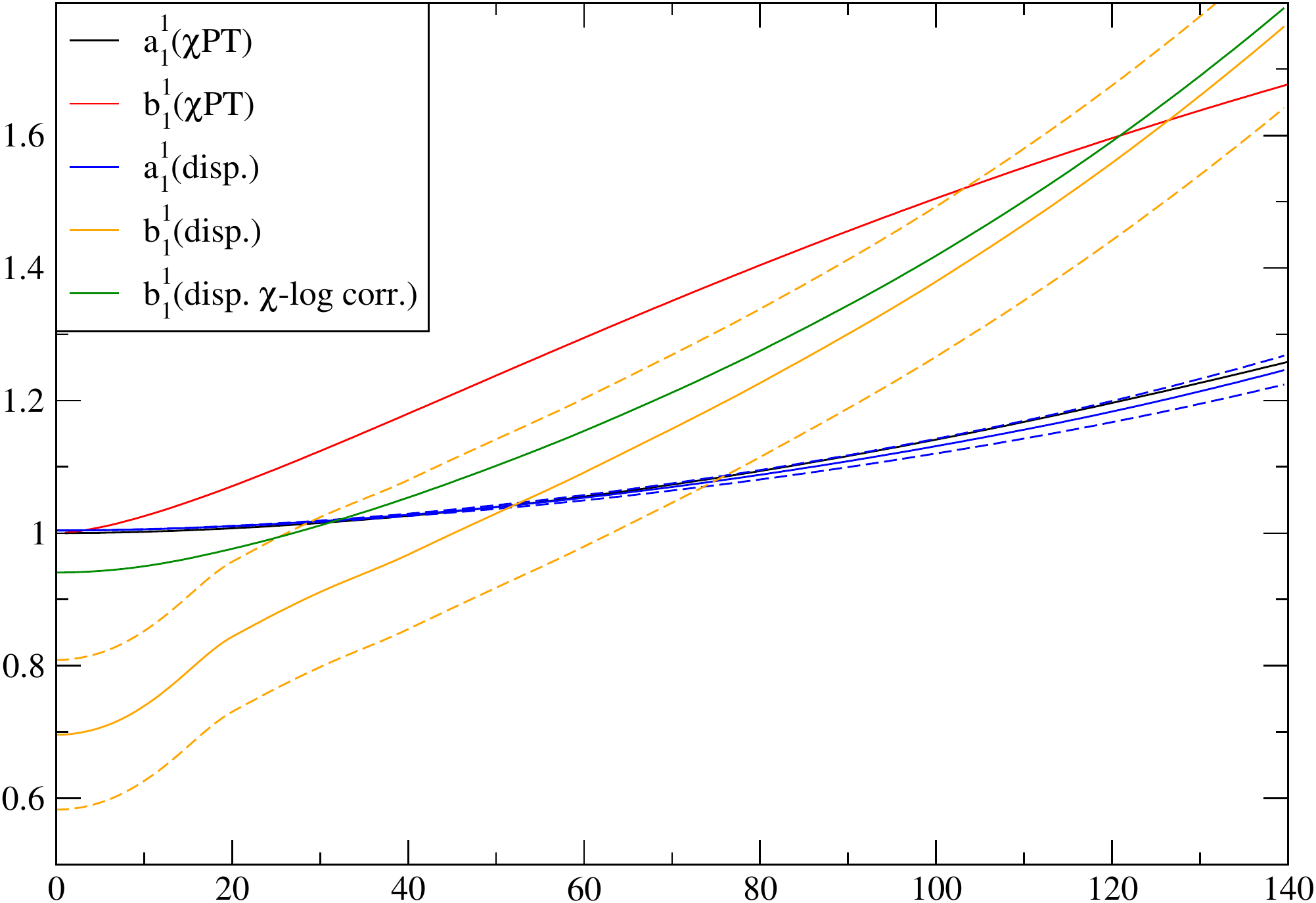}
  \includegraphics[width=8.9cm]{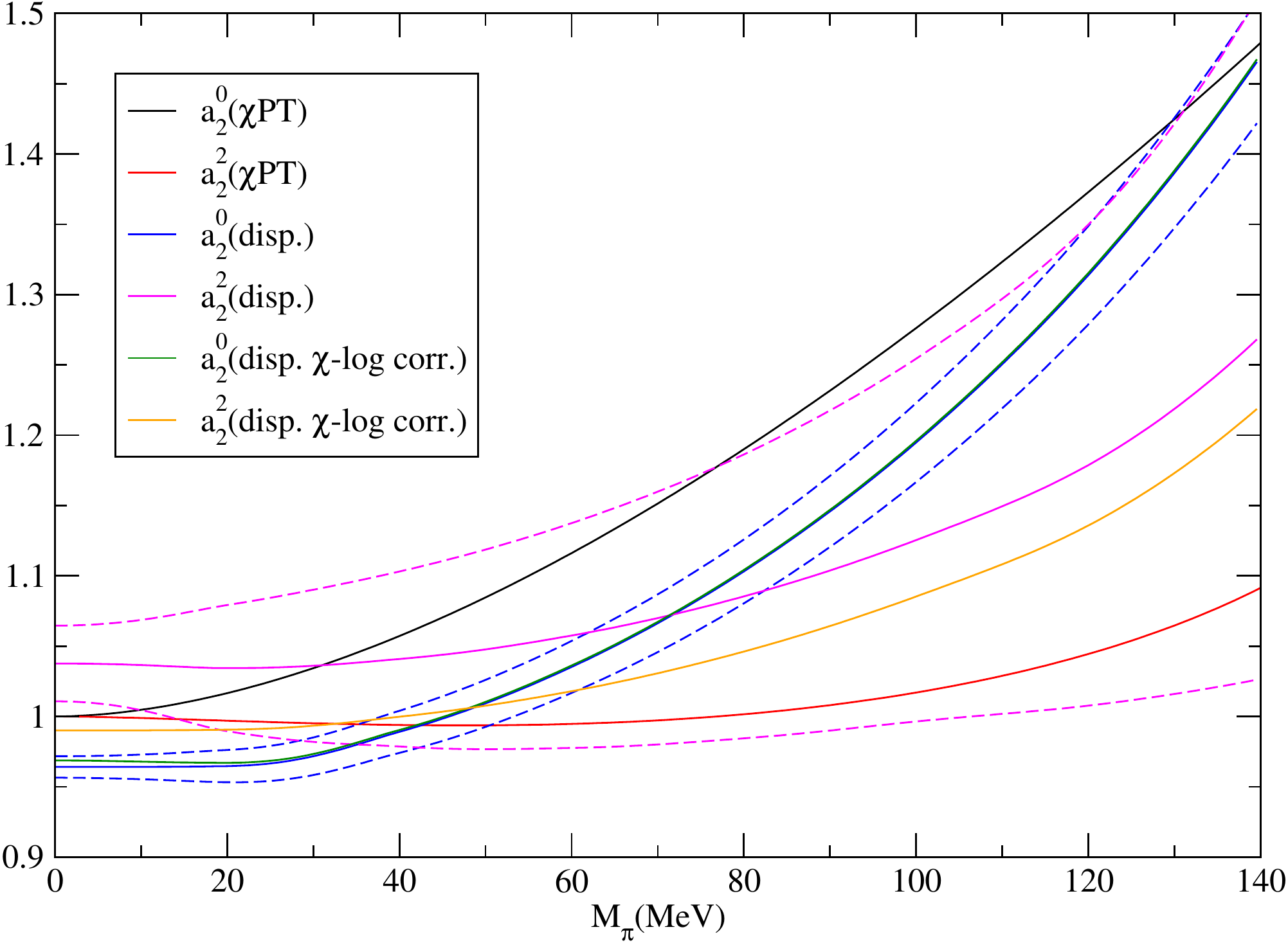} 
  \caption{\label{fig:threshP} Ratios of different threshold parameters to
    the corresponding LO \chpt~formula. The curves labelled \chpt~are the
    ratio of the $\cO(p^6)$ to the LO expressions. Notice that for all
    $S$-wave parameters and $a_1^1$, LO corresponds to $\cO(p^2)$, whereas
    for all the others it is $\cO(p^4)$.}  
\end{figure}
As we move towards the chiral limit, the two-loop \chpt~formulae for the
threshold parameters should provide increasingly accurate values for
them. In the dispersive approach, on the other hand, the only way in which
we have input information about the chiral behaviour is through the
subtraction constants, which then influence the imaginary parts for $4
M_\pi^2 \le s \le s_A$ indirectly, through the solution of the Roy
equations. Comparing the sum rule evaluation of the threshold parameters
for different pion masses to the values obtained with two-loop
\chpt~therefore represents a thorough check of our setup and of the
assumptions we had to make when solving the Roy equations towards the
chiral limit.

This comparison is done in Fig.~\ref{fig:threshP} for the $S$-wave threshold
parameters $b_0^0$ and $b_0^2$, for the $P$-wave parameters $a_1^1$ and
$b_1^1$ and for the two $D$-wave scattering lengths $a^0_2$ and $a^2_2$. To
make the test even more significant we have not plotted the threshold
parameters as such but rather their ratios to the LO \chpt~formula (which
is $\cO(p^2)$ for the first three and  $\cO(p^4)$ for the last three), so that
all of them should go to 1 in the chiral limit.

Overall, the threshold parameters pass this test remarkably well, tracking
the chiral predictions closely and approaching unity in the chiral limit to
within a few percent, with the sole exception of $b_1^1$, which reaches $0.70$
(with an uncertainty of about 0.11, roughly $2.7$ standard deviations below
unity). We discuss this discrepancy in detail below. First, however, we
emphasize that for the parameters that start at $\cO(p^2)$, the ratio to LO
\chpt\ must approach 1 by construction (up to small residual numerical
uncertainties). For all three parameters, the $\cO(p^2)$ contribution is
fixed by the combination $2a_0^0 - 5a_0^2$ taken from \chpt. The genuine test
therefore lies in how the curves evolve away from unity as the pion mass
increases; here, the dispersive and chiral curves follow each other very
closely. For the remaining three parameters, normalizing by the leading
$\cO(p^4)$ contribution divides by an expression involving low-energy
constants (LECs). The test is whether these values are reproduced by the
dispersive integrals over imaginary parts, which start at $\cO(p^4)$ but
resum contributions from all chiral orders. While exact agreement cannot be
expected, the consistency is very satisfactory, particularly considering
that the \chpt\ curves also carry uncertainties (not shown in the figures).

The only prominent discrepancy concerns the chiral limit of $b_1^1$. Its
origin can be understood as follows. The combination of LECs appearing in
$b_1^1$ at $\cO(p^4)$ happens to be scale-independent, which implies that
$b_1^1$ has a vanishing chiral logarithm at $\cO(p^4)$. To reproduce the
correct chiral behaviour, the dispersive sum rule must likewise exhibit a
vanishing chiral log. This is governed by the threshold behaviour of the
integrand, which is proportional to
\begin{equation}
\frac{1}{s^3} \left[2 \Im t_0^0(s)-5 \Im t_0^2(s) - 9 \Im t_1^1(s)\right] -
18 \frac{\Im t_1^1(s)}{(s-4M_\pi^2)^3} \, .
\end{equation}
Near threshold, this integral generates a leading chiral log proportional to
the combination
\begin{equation}
L_1^1:=2 (b_0^0)^2-5 (b_0^2)^2-27(a_1^1)^2 \, .
\end{equation}
At LO in \chpt, this combination vanishes identically, since
$b_0^0(\text{LO})=M_\pi^2/(4\pi F_\pi^2)\stackrel{5\,\text{MeV}}{=}2.700
\cdot 10^{-4}$, $b_0^2(\text{LO})=-M_\pi^2/(8\pi F_\pi^2)\stackrel{5\,
\text{MeV}}{=}-1.350 \cdot 10^{-4}$, and $a_1^1(\text{LO})=M_\pi^2/(24\pi
F_\pi^2)\stackrel{5\,\text{MeV}}{=} 4.501 \cdot 10^{-5}$, which yields
\begin{equation}
L_1^1=\left(\frac{M_\pi^2}{4\pi F_\pi^2}\right)^{2}
\left[2-\frac{5}{4}-\frac{27}{36}\right]=0 \, .
\end{equation}
In the dispersive representation, however, the chiral behaviour of each
threshold parameter is not externally imposed, and their values are
determined solely by the Roy-equation solutions. Expressions for the threshold
parameters in terms of the Chebyshev parameters are readily obtained:
\begin{align}
b_0^0&=a_0^0\left(\frac{4 M_\pi^2}{s^0_R-4
  M_\pi^2}-\frac{(a_0^0)^2}{3}\right)+\frac{4 M_\pi^2}{s_A-4
  M_\pi^2}\sum_{k=0}^8(-1)^k c^0_k \; ,\nonumber \\
a_1^1&=\sum_{k=0}^8(-1)^k c^1_k \; , \qquad \qquad
b_0^2=\frac{4 M_\pi^2}{s_A-4 M_\pi^2}\sum_{k=0}^8(-1)^k c^2_k \; ,
\end{align}
which enables an \textit{a posteriori} check of the values relevant for
$L_1^1$ for the $5\MeV$ central solution. In pion mass units, they are:
\begin{align}
b_0^0(\mathrm{Cheb})&=2.779 \cdot 10^{-4}, \; \; b_0^2(\mathrm{Cheb})=-1.307
\cdot 10^{-4}, \;\; a_1^1(\mathrm{Cheb})=4.451 \cdot 10^{-5}, \nonumber \\
L_1^1(\mathrm{Cheb})&=1.561 \cdot 10^{-8} \; ,
\end{align}
yielding a small nonvanishing chiral log, in contrast to the
\chpt\ requirement. Although the discrepancy between the Chebyshev values
and LO \chpt\ is tiny, 3\% at most, switching from a strictly vanishing
to a nonzero chiral log produces a noticeable effect, pulling $b_1^1$ down
to $\approx 0.7$ in the chiral limit. The chiral
log coefficients of the two $D$-wave scattering lengths are also determined
by the chiral behaviour of the same threshold parameters $b_0^0$, $b_0^2$
and $a_1^1$, and are therefore similarly affected. However, since
their chiral log coefficients are non-zero already, they are only mildly
sensitive to this residual shift: both $a^0_2$ and $a_2^2$ land within 6\%
of unity in the chiral limit, as seen in Fig.~\ref{fig:threshP}. 

As a check that this small discrepancy in the chiral log coefficient is
indeed the cause for this behaviour, we performed the following test: for
the $5\MeV$ solution, the Chebyshev coefficients of the three waves were
slightly adjusted so that $b_0^0$, $b_0^2$ and $a_1^1$ approach the chiral
limit with the coefficient predicted by \chpt. The deviations from the
LO \chpt\ values are quite small, as shown above, and there is some freedom
in how exactly the discrepancy is removed, because all Chebyshev
coefficients appear in the linear combination determining the threshold
parameters. This freedom is exploited to maintain
continuity of the imaginary parts and their first derivatives at $s_A$,
while keeping the relative changes in the Chebyshev coefficients as small as
possible. These shifts are about 4\% in
the $S0$ and $P$ waves and 7.5\% in the $S2$ wave. The impact on the curves
for the imaginary parts is barely visible\footnote{It would, however,
certainly impact the $Q^2$ for the Roy equations. One could in principle
envisage going one step further and solving the Roy equations while
imposing this constraint explicitly, a possibility left for future work.},
yet the resulting curve for $b_1^1$ shifts upward and becomes perfectly
consistent with unity in the chiral limit, as demonstrated in 
Fig.~\ref{fig:threshP}. The chiral behaviour of the two $D$-wave scattering
lengths is likewise affected, but since their deviation from unity was
already small, the corresponding adjustment is much milder, as also shown
in Fig.~\ref{fig:threshP}. 

Finally it is worth examining the sensitivity
of the threshold parameters to the kinematic shift of the $K\bar{K}$
threshold. In the $S$-wave panel of
Fig.~\ref{fig:threshP}, the values of $b_0^0$ and $b_0^2$ obtained from the
Roy equations without any shift of the $K\bar{K}$ threshold are plotted
explicitly; they coincide almost perfectly with the central curves (which
incorporate the shift). The same holds for all other threshold parameters
(these points are omitted from the figures to avoid clutter). This confirms
that while shifting the $K\bar{K}$ threshold is physically well motivated,
the low-energy Roy solutions and threshold parameters are remarkably
insensitive to the precise details of the implementation.

\section{Extension to pion masses above the physical value}
\label{sec:heavierMpi}
\begin{figure}[t]
  \centering
  \includegraphics[width=15cm]{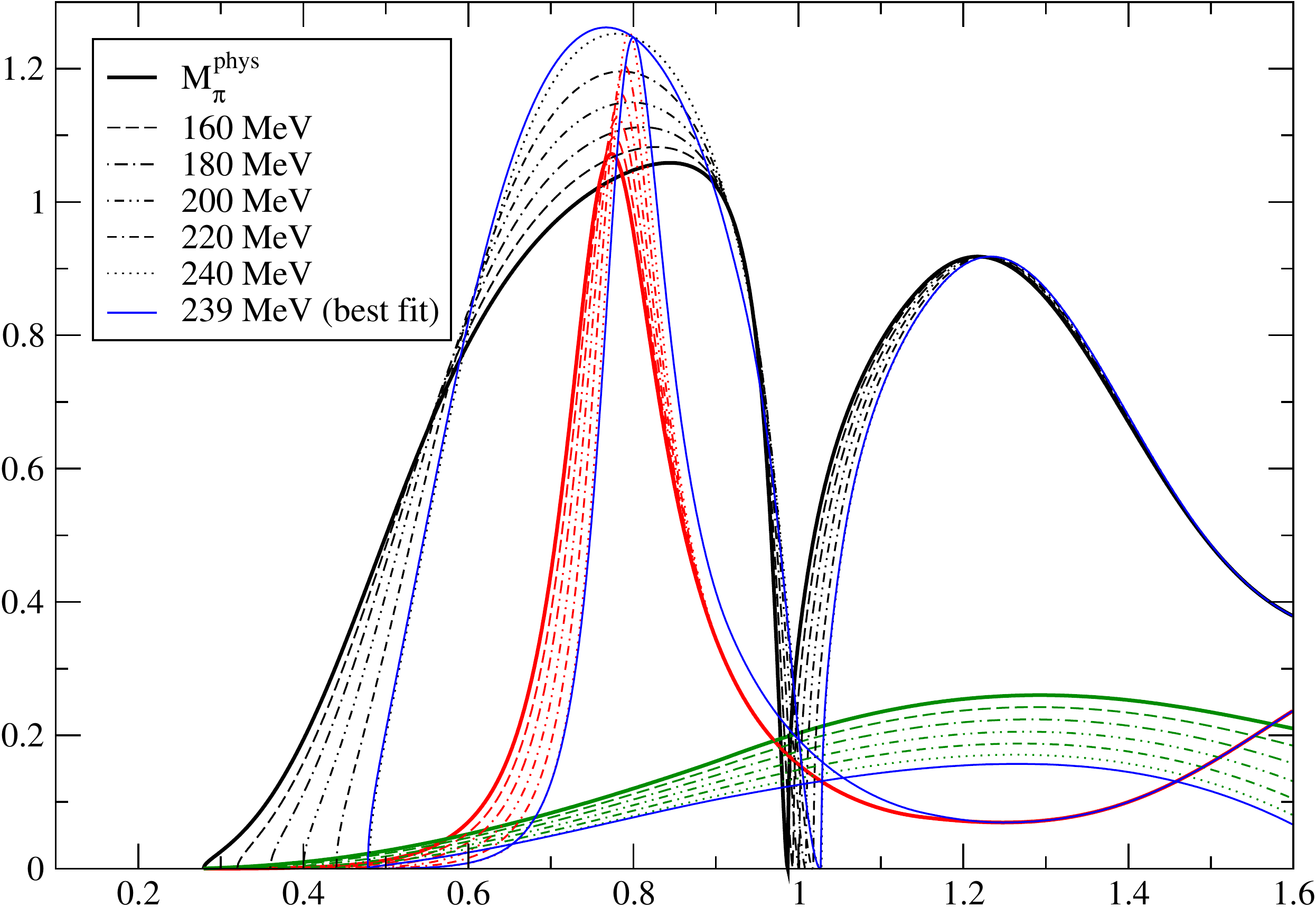}
  \caption{\label{fig:ImtI140240} Imaginary parts below $s_A$ for
    $M_{\pi,\mathrm{phys}} \leq M_\pi \leq 240\MeV$ evaluated for our
    central input. The curve with $M_\pi=239\MeV$ labeled as (best fit) has
    been evaluated for input parameters $x_{It0}=-0.027$ and
    $\Delta \Im t_1^1(s_A)=0.060$, obtained from a fit to the lattice
    data in~\cite{Rodas:2023nec}, as discussed below.}  
\end{figure}
As discussed above, if we solve the Roy equations for pion masses higher
than physical, at some value we are going to face difficulties related to
the convergence of the chiral series for the two scattering lengths and to
the assumption that the input above $s_A$ remains constant with respect to
the pion mass. Concerning the first issue, Fig.~11
in~\cite{Colangelo:2001df} provides an indication that values below
$200\MeV$ should be safe and that somewhere between 200 and $300\MeV$ the
chiral $\cO(p^6)$ prediction for $a_0^0$ should not be reliable anymore. As
for the second issue, only lattice calculations can provide reliable
information on the $M_\pi$ breakdown value. Since there are lattice data on
the three lowest partial waves for $M_\pi=239\MeV$~\cite{Rodas:2023nec}, we
evaluate numerically Roy-equation solutions within the same framework
adopted for $M_\pi<M_{\pi,\mathrm{phys}}$ ({\em i.e.} same input above
$s_A$ as for $M_{\pi,\mathrm{phys}}$ other than for the shifted $K\bar{K}$
threshold) for $M_{\pi,\mathrm{phys}} < M_\pi \leq 240\MeV$. Our goal is to
make a first exploration of this region with the approach developed here
and to learn, through a comparison to the lattice data at $M_\pi=239\MeV$, 
whether the difficulties mentioned above show up already for this value of
the pion mass.

The first conclusion of this analysis is that solving the Roy equations for
$M_\pi>M_{\pi,\mathrm{phys}}$ with the same input adopted for
$M_\pi<M_{\pi,\mathrm{phys}}$ is completely unproblematic. The quality of
the solution slightly improves for $M_\pi=160\MeV$ ($Q^2=4.4\cdot
10^{-11}$) and then slowly increases, reaching $Q^2=1.3\cdot 10^{-9}$ for
$M_\pi=240\MeV$. The imaginary parts of the three lowest waves are shown in
Fig.~\ref{fig:ImtI140240} for $M_\pi$ between the physical value and
$240\MeV$ in $20\MeV$ intervals. Even in this region the dependence on
$M_\pi$ remains smooth, and indeed the explicit solutions shown here
are not significantly different from extrapolations obtained from the
solutions with $M_\pi < M_{\pi,\mathrm{phys}}$.

\subsection{Comparison and fit to lattice data for $M_\pi=239$ MeV}
\begin{figure}[th]
  \centering
  \includegraphics[width=9.3cm]{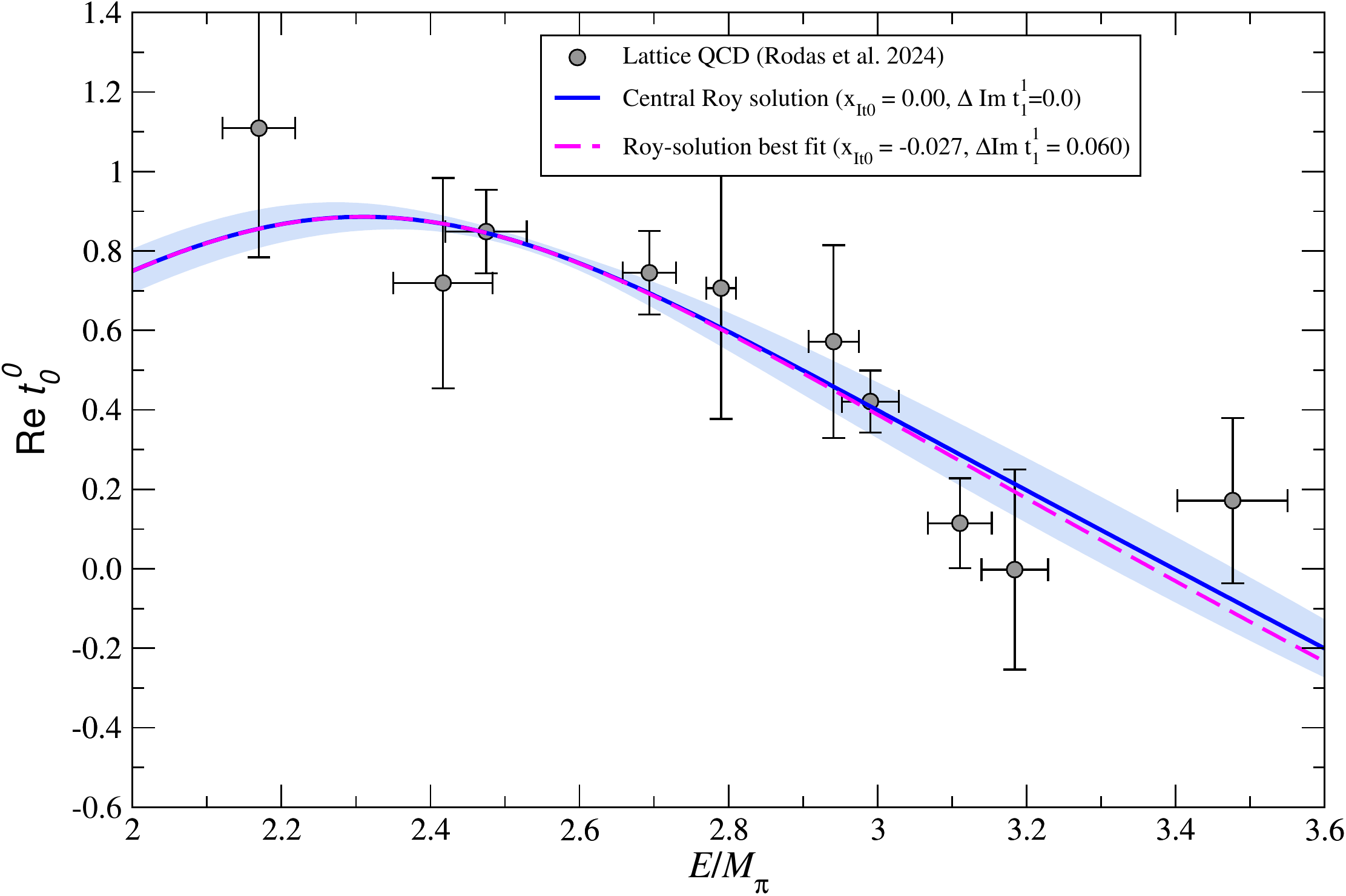}
  \includegraphics[width=9.3cm]{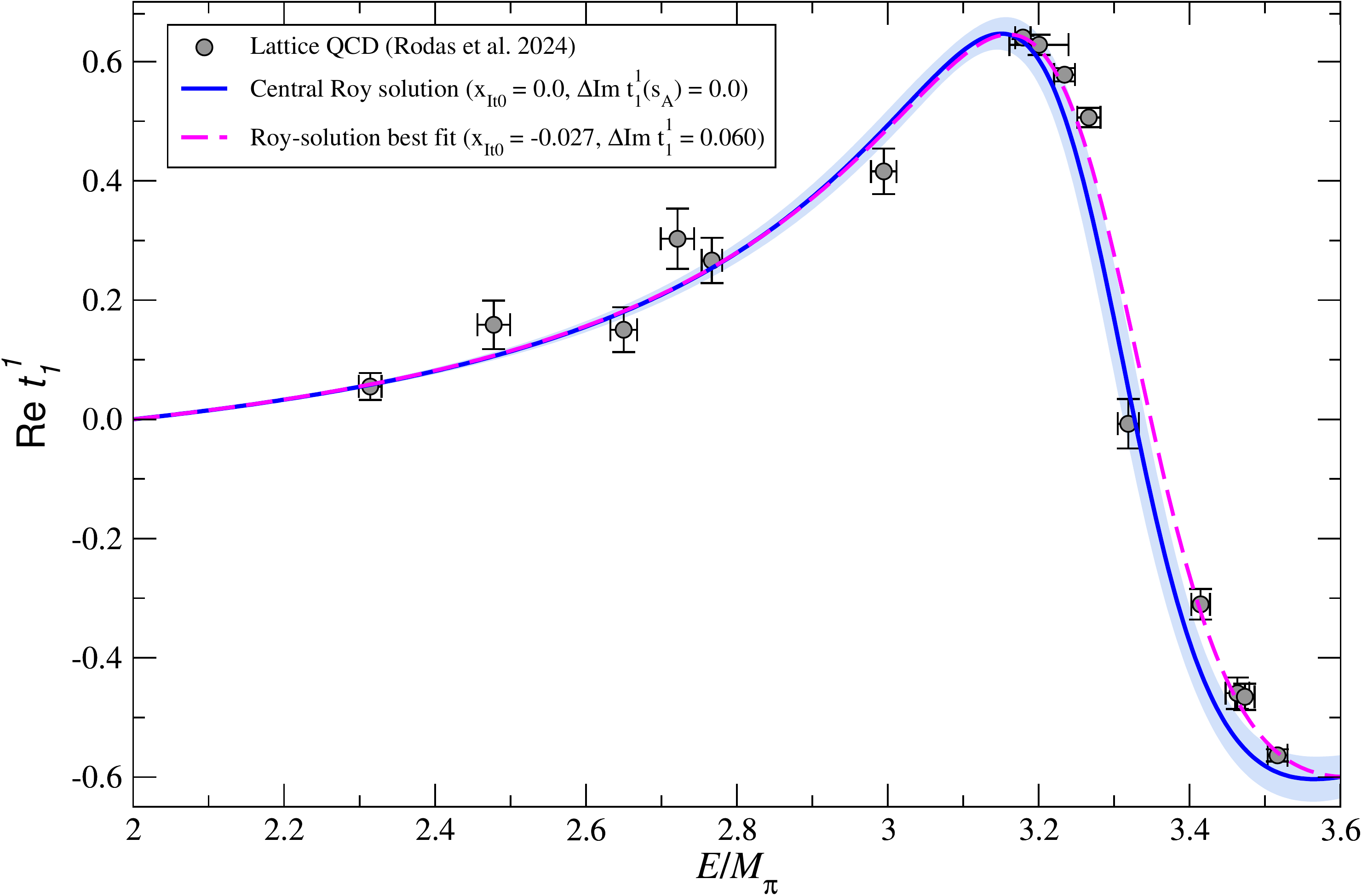}
  \includegraphics[width=9.3cm]{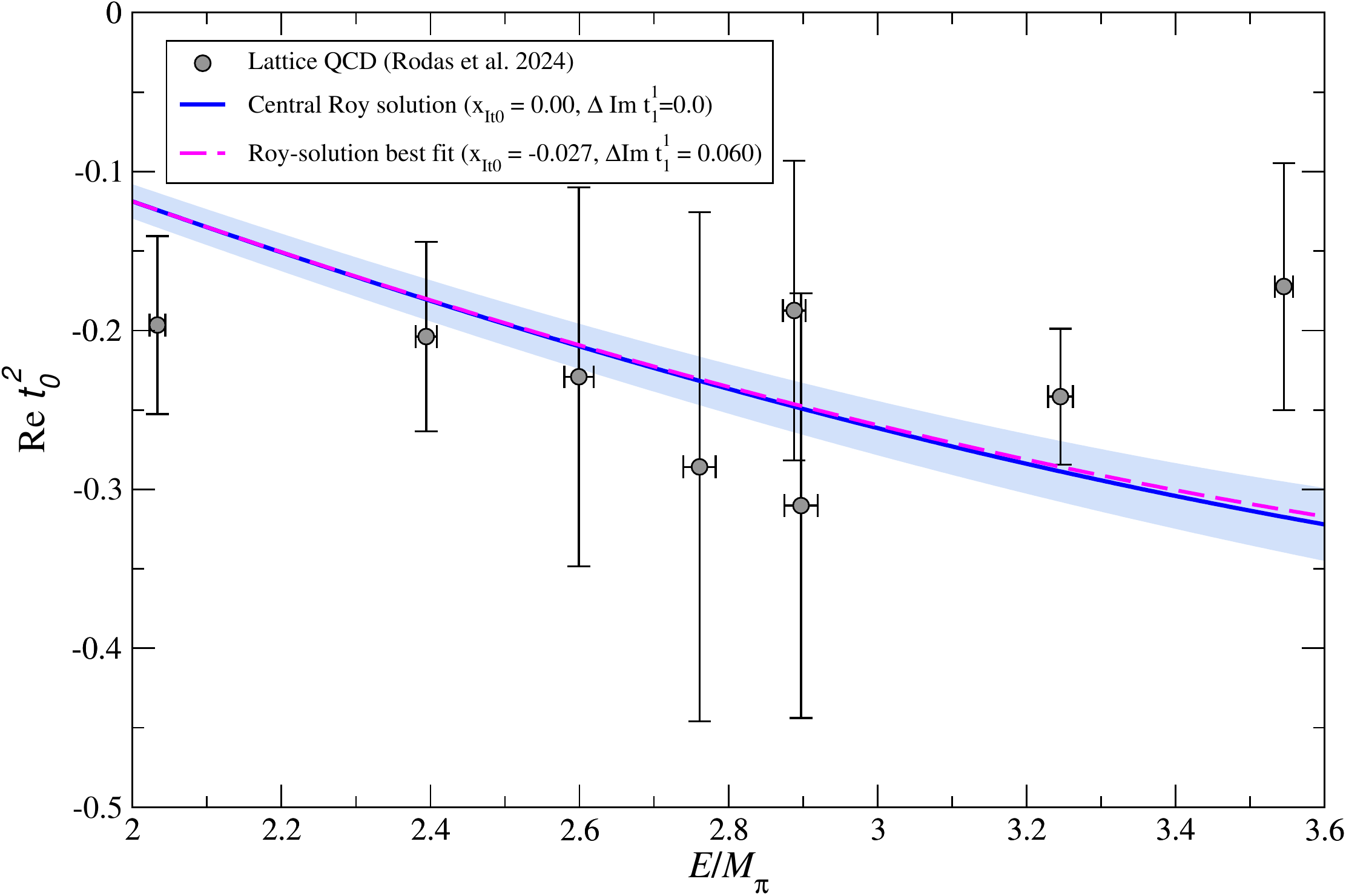} 
  \caption{\label{fig:CompRodas} Real parts of the three lowest partial
    waves evaluated for $M_\pi=239\MeV$. The continuous curves and bands
    show the Roy solutions obtained with our central
    input ($x_{It0}=0$ and $\Delta \Im t_1^1(s_A)=0$) and the corresponding
    uncertainties, whereas the dashed magenta curves represent our best
    fit to the lattice data in~\cite{Rodas:2023nec}, which are also shown.}  
\end{figure}
Lattice data on the lowest partial waves of the $\pi \pi$ scattering
amplitude for a pion mass of $239\MeV$ have been provided and then analysed
in a series of papers by the Hadron Spectrum
Collaboration~\cite{Wilson:2015dqa,Briceno:2016mjc,
  Briceno:2017qmb,Rodas:2023gma,Rodas:2023nec}. In the analysis
paper~\cite{Rodas:2023nec} lattice data have been fit with a representation
which has later been used as input in the Roy integrals, to check whether
the output was compatible with the input. In the end the best fit to the
data which at the same time represented an acceptable solution to the Roy
equations was presented as the best outcome of the analysis. In the
present approach, if we stick to our central input for the imaginary parts
for $s \geq s_A$ and the two scattering lengths, we do not have any fit
parameters and can simply calculate the real parts from our Roy solutions
and compare them to the lattice data. This is done in
Fig.~\ref{fig:CompRodas} which shows the real parts of our central Roy
solutions (after tuning the pion mass to exactly $239\MeV$) and the lattice
data in~\cite{Rodas:2023nec}. The error bands have been calculated
following the same recipe discussed in Sect.~\ref{sec:Unc} for the region
$M_\pi < M_{\pi,\mathrm{phys}}$. Overall we observe good agreement, in
particular for the two $S$ waves. For the $P$ wave, both because of the
small error bars of the lattice data and the strong $s$-dependence in the
resonance region, some local discrepancies between the Roy-solution curves
and the data are visible. 
\begin{table}[tbp]
  \begin{tabular}{l c cc cc c}
    \toprule
    \multirow{2}{*}{} & \multirow{2}{*}{$N_\mathrm{pts}$} 
    & \multicolumn{2}{c}{Central Roy Solution} 
    & \multicolumn{2}{c}{Best-Fit Roy Solution} 
    & \multirow{2}{*}{$\Delta \chi^2$} \\
    \cmidrule(lr){3-4} \cmidrule(lr){5-6}
    & & $\chi^2$ & $\chi^2 / \mathrm{dof}$ & $\chi^2$ & $\chi^2 / \mathrm{dof}$ & \\
    \midrule
    $S0$ wave & 10 & $4.67$  & $0.47$ & $4.67$  & $0.52$ & $-0.00$ \\
    $P$ wave & 15 & $18.25$ & $1.22$ & $9.23$  & $0.66$ & $-9.02$ \\
    $S2$ wave &  8 & $6.75$  & $0.84$ & $6.44$  & $0.80$ & $-0.31$ \\
    \midrule
    \text{Total} & 33 & $29.7$ & $0.90$ & $20.3$ & $0.66$ & $-9.3$ \\
    \bottomrule
  \end{tabular}
  \centering
  \caption{$\chi^2$ values of the two Roy-equation solutions at $M_\pi =
    239\MeV$ with respect to the lattice QCD data of Rodas
    \textit{et al.}~\cite{Rodas:2023nec}. The uncertainties are combined in
    quadrature as $\sigma_{\mathrm{tot}, i}^2 = [\sigma(\mathrm{Re}\,t_i)]^2 +
    \left[\frac{d\,\mathrm{Re}\,t}{d(E/M_\pi)} \cdot \sigma(E/M_\pi)_i\right]^2
    + \sigma_{\mathrm{th}}^2(E_i)$, where $\sigma_{\mathrm{th}}$ is the
    theoretical Roy uncertainty band.} 
  \label{tab:chi2_comparison}
\end{table}
To be more quantitative, we have evaluated the $\chi^2$ of the three
curves with respect to the lattice data, taking into account the lattice
uncertainty both in $\Re t^I_\ell$ and in $E/M_\pi$ as well as the
theoretical uncertainty. The numbers shown in
Table~\ref{tab:chi2_comparison} confirm the very good agreement for the two
$S$ waves and provide a measure of the level of disagreement for the $P$ wave,
which is only mild (the $p$-value corresponding to $\chi^2=18.25$ for $15$
data points is equal to $25.3\%$). We stress that this is obtained with no
free parameters and therefore represents a further confirmation of the
robustness of the present approach and of the correctness of the
assumptions involved. 

With the availability of these lattice data we can, however, relax some of
the assumptions, like those about the input for $s \geq s_A$, and let the
data tell us how this input should be adapted to the value of $M_\pi$
considered. In particular, since the values of the imaginary parts at $s_A$
are much more relevant than the exact shape higher up in energy, we will do
the following: evaluate the $\chi^2$ with respect to the data for Roy
solutions obtained with different values of $x_{It0}$ (which fixes $\Im
t_0^0(s_A)$) and of $\Delta \Im t_1^1:= \Im t_1^1(s_A) - \Im
\overline{t_1^1(s_A)}$ and then minimize the $\chi^2$ with respect to these
two parameters. The minimum has been found at $x_{It0}=-0.027$ and $\Delta
\Im t_1^1=0.060$ and corresponds to a total $\chi^2=20.3$ with a decrease
of $9.3$, almost completely due to the $P$ wave, as shown in
Table~\ref{tab:chi2_comparison}. The improvement in the agreement with the
lattice data is evident by looking at Fig.~\ref{fig:CompRodas}, which also
makes it clear that it is the highest four lattice data for the $P$ wave
which pulled the Roy-solution curve to the right, a result obtained by
shifting $\Im t_1^1(s_A)$ up by $0.060$. This shift is clearly visible in
Fig.~\ref{fig:ImtI140240}, whereas the one in the $S0$ wave is quite small.
We stress that the uncertainties in these two parameters evaluated
according to the recipe discussed in Sect.~\ref{sec:Unc} are
$\sigma(x_{It0})=0.045$ and $ \sigma(\Delta \Im t_1^1)=0.0225$, which shows
that the lattice data require less than a $1 \sigma$ shift in $x_{It0}$ and
a $2.7 \sigma$ shift in $ \Delta \Im t_1^1$. The latter is not a cause for
concern, since the uncertainty estimate was specifically designed for the
region $M_\pi < M_{\pi,\mathrm{phys}}$ and then adopted as such also to the
$M_\pi > M_{\pi,\mathrm{phys}}$ region, but clearly indicates that here the
guidance of lattice data is much more reliable than a priori
assumptions.

Overall, this comparison shows that analysing lattice data on the basis of
the $M_\pi$-dependent Roy solutions is not only possible but indeed very
promising. This discussion is only a first step in this direction, and a
proof of principle that shows that this approach should be pursued more
systematically in the future. More lattice data even at higher masses
should be analysed (starting from the $283\MeV$ data also presented
in~\cite{Rodas:2023nec}), but this will require a thorough analysis of the
limitations of the present approach as one goes to even higher pion
masses. In particular, the assumption that the phenomenological input above
$s_A$ can be adopted without changes can no longer be justified. The way
forward is to rely on lattice data for that input too, as was pioneered in
Ref.~\cite{Cao:2023ntr}. We stress that this latter paper is mainly focused
on the case in which the pion mass is so high that the $\sigma$ becomes
stable (which happens around $M_\pi=300\MeV$), which is significantly
higher than the range considered here. Indeed, if the $\sigma$ becomes
stable the Roy equations have to be reformulated to take into account its
pole on the real axis, as done in Ref.~\cite{Cao:2023ntr}. Another
difference with respect to the present setup is that, in that region,
\chpt\ cannot reliably predict the $S0$-wave scattering length anymore, so
that possibly even that input may have to be taken from the lattice.

Alternatively, one could try to circumvent the problem of the chiral
convergence of $a_0^0$ in the following way: in
Ref.~\cite{Colangelo:2001df} it has been argued that instead of relying on
the two scattering lengths as input, one can make a matching between the
dispersive and the chiral representations below threshold, which
dramatically improves the chiral convergence of the subtraction
constants. If one is interested in studying Roy-equation solutions for
masses higher than physical and even beyond the relatively low pion mass
value of $239\MeV$ we just discussed, this approach offers clear advantages
and should be adopted in future work in this direction.

\section{Conclusions and Outlook}
\label{sec:concl}
The analysis presented in this paper shows that it is possible, on the
basis of a few very reasonable assumptions, to solve the Roy equations for
pion masses below the physical value, as well as above it. The main
assumption is that the high-energy input to the Roy equations is
insensitive to the exact value of the light-quark masses and would remain
essentially unchanged in the chiral limit. While such an expectation sounds
very plausible, the precise implementation requires some care.  In the
present analysis, we chose to solve the Roy equations below
$s_A=0.84\GeV^2$ ($\sqrt{s_A}=0.917\GeV$). Admittedly, light-quark mass
effects will not be completely negligible at such energies and our analysis
has shown that the Roy equations themselves require at least some
flexibility in the input at and above $s_A$ as the pion mass changes. We
concluded, however, that as the light-quark masses are reduced to zero, the
phenomenological input does not need to change beyond its
uncertainties. Within this framework it has been possible to obtain
Roy-equation solutions of good quality all the way down to pion masses of
$5\MeV$, essentially in the chiral limit.

The reliability of these results has been verified by evaluating several
threshold parameters with the help of Wanders' sum rules. These contain a
contribution proportional to $2a_0^0-5 a_0^2$ (present only for $b^{0,2}_0$
and $a_1^1$) plus integrals over imaginary parts of partial waves. As the
pion mass is reduced, the term proportional to $2a_0^0-5 a_0^2$ reproduces
by construction the correct chiral behaviour, but the integrals change both
because the lower limit of integration (the $\pi \pi$ threshold) shifts
towards zero, and because the shape of the imaginary parts changes as
dictated by the Roy equations. Evaluating these sum rules for varying pion
masses and comparing them to the chiral predictions for the threshold
parameters represents a test of the mass dependence of these integrals. We
find that the behaviour of all resulting threshold parameters closely
follows the predictions of \chpt. This demonstrates that the
underlying assumptions are robust and that the light-quark-mass dependence
obtained here for the $\pi \pi$ scattering amplitude in the region below
$s_A$ is reliable.

The main motivation for this analysis lies in the possibility of
establishing a closer connection with other approaches in which the pion
mass (or the light-quark masses) can naturally be varied, such as lattice
QCD~\cite{Beane:2007xs,NPLQCD:2011htk,Feng:2009ij,Dudek:2012gj,Feng:2010es,
  Dudek:2013yja,Wilson:2015dqa,Bali:2015gji,Bulava:2016mks,Culver:2019qtx,Fischer:2020jzp,
  Briceno:2016mjc,Briceno:2017qmb} (based on the finite-volume methodology
of Ref.~\cite{Luscher:1990ux}, see Ref.~\cite{Briceno:2017max} for a
review) or the S-matrix
bootstrap~\cite{Guerrieri:2018uew,Guerrieri:2020bto,Albert:2022oes,
  Tourkine:2023xtu,Guerrieri:2024jkn}. Without relying on lattice data
themselves as input at a given unphysical mass, such a connection was
previously only possible for physical values of the quark masses. This
significantly reduces the usefulness of such comparisons, and moreover
leaves most of the burden on the other approaches, for which reaching the
physical pion mass can be very costly. The main novelty of the present
analysis is to prove that it is indeed possible to vary the light-quark
masses within a dispersive approach, without significant loss of rigour. It
is of course well known that \chpt~represents a rigorous tool to control 
the dependence on quark masses, but it suffers from two main limitations:
first, even if it can predict the analytic form of the light-quark-mass
dependence, it still requires the determination of the relevant LECs to
make numerical predictions, and those which are responsible for the pure
quark-mass dependence are difficult (and in some cases impossible) to
obtain from phenomenology; second, the region of validity of \chpt~is
limited, whereas a dispersive approach can go well beyond that region.

Having established the validity of the approach for pion masses below the
physical value, we have also explored the region above. We have shown that,
working under the same assumptions, it is possible to obtain Roy-equation
solutions of excellent quality up to pion masses of about $240\MeV$. A
comparison to existing lattice data~\cite{Rodas:2023nec} for the three
lowest partial waves at $M_\pi=239\MeV$ has shown good agreement even
without adapting or introducing any free parameters in our framework. The
assumption about strict independence of the input above $s_A$ from the pion
mass can be relaxed, however. As we have shown, one can describe such a
dependence in terms of a few parameters and use these to fit the lattice
data. This further improves the already good agreement of our Roy
solutions with the lattice data. This proof of principle paves the way to a
more systematic use of Roy-equation solutions in analysing lattice data for
the $\pi\pi$ scattering amplitude at pion masses above the physical value,
potentially even beyond the conservative limit of $240\MeV$ adopted
here. Such a region of pion masses, even up to where the $\sigma$ becomes
stable, was studied with Roy equations in Ref.~\cite{Cao:2023ntr}, but with
a different implementation than the one adopted here (see also
Refs.~\cite{Cao:2024zuy,Cao:2025hqm} for studies of the Roy-Steiner
equations for $\pi K$ scattering at unphysical quark masses to interpret
lattice data in that channel).

Among the observables of particular interest are the pole positions of the
lightest resonances. With the solutions of the Roy equations at hand it is
straightforward to calculate these without any significant systematic
uncertainties related to the analytic continuation into the complex plane,
as was shown in~\cite{Caprini:2005zr}. The same is true if these solutions
are obtained for unphysical pion masses, which opens the way to
applying the formalism developed here to analyse lattice data on the
partial waves in order to determine the pole positions. A thorough analysis
of the quark mass dependence of the $\rho$ and $\sigma$ resonance
parameters (for previous work
see~\cite{Hanhart:2008mx,Molina:2020qpw,Niehus:2020gmf,Cao:2023ntr,Rodas:2023nec,Pelaez:2026fmu})
is one of the most immediate applications of the approach developed here and
will be presented in a forthcoming publication~\cite{Colangelo-Zwicky}.

\section*{Acknowledgements} I thank Stephan Thurnherr who participated in the
early stages of this project working on his master thesis, and Roman Zwicky
for pushing me to carry out such an analysis, and for several useful
discussions. I thank Martin Hoferichter, Peter Stoffer and Roman Zwicky
for a careful reading and useful comments on the manuscript. Useful
discussions with Andrea Guerrieri are also gratefully acknowledged. Most of
the work presented here was performed during a sabbatical semester spent at
the CERN Department of Theoretical Physics. I warmly thank them for the
hospitality and stimulating environment. This project has benefitted from
the support of AI, Gemini in particular. The latter carried out very
quickly and efficiently many time-consuming tasks related to the numerical
work presented here, mostly with the help of Python scripts, which I would
not have been able to write---certainly not so fast. It also helped
with proofreading the manuscript.

\appendix

\section{Numerical values of input parameters and driving terms}
\label{app:nums}
In the calculations described above we have made use of the following input
parameters:
\begin{equation}
M_{\pi,\text{phys}}=139.57018\MeV\, , \quad F_{\pi,\text{phys}}=92.07\MeV,
\quad M_{K^+}=493.677\MeV
\, .
\end{equation}
The value of the pion mass has not been adapted to the latest
PDG~\cite{ParticleDataGroup:2026mpi} value, in order 
to keep full compatibility with the earlier Roy solutions which have been
used
here~\cite{Ananthanarayan:2000ht,Colangelo:2001df,Caprini:2011ky}. Since
the shift is minute ($139.57039\MeV$), the change would probably be
invisible. 

In the SU(2) \chpt~calculations which we made use of here, we have used as
input the values of the LECs taken from Ref.~\cite{Colangelo:2001df}, which
we report here for convenience. First the $\cO(p^4)$ LECs:
\begin{align}
\bar{\ell}_1&=-0.4(6) \, , \; \; \bar{\ell}_2=4.3(1) \, , \; \; \bar{\ell}_3=2.9(2.4) \, , \;
\; \bar{\ell}_4=4.4(2) \, ,
\end{align}
and then the $\cO(p^6)$ LECs ($\tilde{r}_i=(16 \pi^2)^2r_i^r(M_\rho^2)$):
\begin{align}
\tilde{r}_1&=-1.5 \,, \; \; \tilde{r}_2=3.2 \,, \; \; \tilde{r}_3=-4.2 \,, \; \;
\tilde{r}_4=-2.5 \,, \; \; \tilde{r}_5=3.8 \,, \; \; \tilde{r}_6=1.0 \, .
\end{align}

The quark mass dependence of $F_\pi$ is known in \chpt~to
$\cO(p^6)$~\cite{Bijnens:1998fm}, and has been calculated here 
with the following numerical values of the relevant input parameters:
\begin{equation}
F=85.8\MeV \, , \; \; \tilde{r}_F(M_\rho^2)=0 \;,
\end{equation}
where $\tilde{r}_F$ is the combination of $\cO(p^6)$ LECs which appears in
$F_\pi$, according to the representation in Ref.~\cite{Colangelo:2003hf}.

For the light-quark mass dependence of $M_K$ shown in Fig.~\ref{fig:MKMpi},
we evaluated the NLO SU(3) \chpt\ mass formulas at the renormalization
scale $\mu=M_\rho=0.77\GeV$. The relevant $\cO(p^4)$ LECs were taken from the
FLAG 2021 $N_f=2+1+1$ review~\cite{FlavourLatticeAveragingGroupFLAG:2021npn}:
\begin{align}
L_4^r&=0.09 \cdot 10^{-3} \, , \; \; L_5^r=1.19 \cdot 10^{-3} \, , \; \;
L_6^r=0.16 \cdot 10^{-3} \, , \; \; L_8^r=0.55 \cdot 10^{-3} \, .
\end{align}
The leading-order strange-quark mass parameter $m_s^2$ and the
scale-independent LEC $L_7$ were determined by requiring that the NLO expressions
reproduce the physical masses $M_{K^+}=493.677\MeV$ and $M_\eta=547.3\MeV$ at the
physical pion mass, yielding $m_s^2=0.2264\GeV^2$ and $L_7=-0.24 \cdot 10^{-3}$.

The driving terms are taken over from the analysis in
Ref.~\cite{Caprini:2011ky}, and are expressed as simple polynomials in
$s$, after factoring out the $s- 4 M_\pi^2$ factor dictating the correct
threshold behaviour:
\begin{equation}
d^I_\ell(s)=(s-4M_\pi^2) \sum_{n=0}^3 \kappa^I_{n} s^n \, .
\end{equation}
The values we have used in the present analysis are given in Table~\ref{tab:driving_terms}.
The estimated uncertainties are at the level of 8\%, 40\%, and 20\% for the $S0$,
$P$, and $S2$ waves respectively. The large relative uncertainty in the $P$
wave is a consequence of the fact that the driving term in that channel is
very small.
\begin{table}[tbhp]
  \centering
  \small
  \setlength{\tabcolsep}{6pt}
  \begin{tabular}{crrrr}
  \toprule
  Channel & $\kappa^I_0\;[\text{GeV}^{-2}]$ & $\kappa^I_1\;[\text{GeV}^{-4}]$ & $\kappa^I_2\;[\text{GeV}^{-6}]$ & $\kappa^I_3\;[\text{GeV}^{-8}]$ \\
  \midrule
  $S0$ &  $0.008399$ & $0.326592$ & $-0.071785$ &  $0.015655$ \\
  $P$  & $-0.000763$ & $0.016173$ &  $0.006180$ & $-0.000737$ \\
  $S2$ &  $0.004631$ & $0.147164$ & $-0.105116$ &  $0.024806$ \\
  \bottomrule
  \end{tabular}
  \caption{Coefficients of the polynomial expansion for the driving terms.}
  \label{tab:driving_terms}
\end{table}

\section{Specification of the $M_\pi$-dependent Roy solutions}

For $s \le s_A$ the Roy-equation solutions are expressed in terms of 9
Chebyshev parameters for each wave, plus the two values $s^I_R$ of $s$
where the $S0$ and $P$ phases go through $\pi/2$. The precise form of the
parametrization is given above, see Eq.~\eqref{eq:ChebI}. Here we provide the
values of these parameters for the central solution and a polynomial
parametrization of their $M_\pi$-dependence.

The pion-mass dependence of each parameter $c^I_k$ or $s^I_R$ is
parametrized by a quadratic polynomial in the normalized mass squared
variable $y_\pi := (M_\pi / M_{\pi,\text{phys}})^2$ (with
$M_{\pi,\text{phys}} = 139.57018\,\text{MeV}$):  
\begin{equation}
c^I_k(y_\pi) = c^I_{k,0} + c^I_{k,1} y_\pi + c^I_{k,2} y_\pi^2 \, ,
\end{equation}
and analogously for $s^I_R$.  For the $P$ wave, the Chebyshev coefficients
$\tilde{c}_n^1 \equiv c_n^1 / (4 M_\pi^2)$ are normalized by $4 M_\pi^2$ so
that they possess a smooth chiral limit, and the expansion in $y_\pi$
applies to them. Over the entire range $M_\pi \in [5, 139.57]\,\text{MeV}$
and energies $s \le s_A$, this quadratic parametrization reproduces the
exact numerical solutions of the Roy equations with an accuracy better than
$0.03^\circ$ in the $S0$ phase shift, and better than $0.005^\circ$ in the
$P$ and $S2$ phase shifts (corresponding to an amplitude difference $\Delta
\text{Re}\,t_\ell^I \le 5 \cdot 10^{-4}$), which is completely negligible
compared to the systematic uncertainties of the dispersive solutions.

The values of the expansion coefficients $c^I_{k,n}$ as determined in the
present analysis are provided in Table~\ref{tab:chebyshev_coeffs}. For
convenience, the table also lists the exact values of the parameters at the
physical pion mass, $c^I_k(\text{phys})$ and $s_R^I(\text{phys})$, obtained
from the central Roy solution at $M_{\pi,\text{phys}} = 139.57018\MeV$ as
well as for the central solution and the best fit to the lattice data at 
$M_\pi = 239\,\MeV$.

\begin{table}[!htbp]
  \centering
\small
\setlength{\tabcolsep}{5pt}
\begin{tabular}{c l rrrrrr}
\toprule
Wave & Parameter & $c^I_{k,0}$ & $c^I_{k,1}$ & $c^I_{k,2}$ & $c^I_k(\text{phys})$ & $c^I_k(239\,\text{C})$ & $c^I_k(239\,\text{F})$ \\
\midrule
\multirow{10}{*}{\rotatebox{90}{$S0$-wave}} & $c_{0}^0$ & $1.57282$ & $0.18321$ & $-0.11123$ & $1.64543$ & $1.28735$ & $1.25523$ \\
& $c_{1}^0$ & $-1.10406$ & $0.00825$ & $0.02479$ & $-1.07188$ & $-0.91270$ & $-0.89258$ \\
& $c_{2}^0 \cdot 10$ & $-1.74426$ & $0.54394$ & $0.11421$ & $-1.08610$ & $0.09853$ & $0.26047$ \\
& $c_{3}^0 \cdot 10$ & $1.02242$ & $-0.94199$ & $0.32425$ & $0.40119$ & $0.04624$ & $0.03247$ \\
& $c_{4}^0 \cdot 10^2$ & $-4.32356$ & $4.24680$ & $-1.85556$ & $-1.90162$ & $-0.61373$ & $-0.61921$ \\
& $c_{5}^0 \cdot 10^2$ & $1.24851$ & $-1.98022$ & $1.00770$ & $0.25795$ & $0.06407$ & $0.07687$ \\
& $c_{6}^0 \cdot 10^3$ & $-5.13494$ & $8.92838$ & $-4.85399$ & $-0.95975$ & $-0.07093$ & $-0.06074$ \\
& $c_{7}^0 \cdot 10^3$ & $1.65438$ & $-3.55653$ & $2.05665$ & $0.11001$ & $0.03853$ & $0.05164$ \\
& $c_{8}^0 \cdot 10^3$ & $-0.78551$ & $1.53081$ & $-0.85143$ & $-0.08733$ & $0.05443$ & $0.07114$ \\
& $s_R^0\,[\text{GeV}^2]$ & $0.80682$ & $-0.08585$ & $0.00831$ & $0.72935$ & $0.65961$ & $0.64875$ \\
\midrule
\multirow{10}{*}{\rotatebox{90}{$P$-wave}} & $\tilde{c}_{0}^1$ & $0.40763$ & $0.03698$ & $0.00778$ & $0.45243$ & $0.59477$ & $0.61553$ \\
& $\tilde{c}_{1}^1 \cdot 10$ & $-0.50280$ & $0.04359$ & $0.02521$ & $-0.43442$ & $-0.15572$ & $0.01185$ \\
& $\tilde{c}_{2}^1 \cdot 10^2$ & $-1.30897$ & $0.49815$ & $-0.15725$ & $-0.96421$ & $-0.65528$ & $-0.46999$ \\
& $\tilde{c}_{3}^1 \cdot 10^4$ & $2.46688$ & $-2.26981$ & $4.20567$ & $4.18740$ & $10.15146$ & $12.43390$ \\
& $\tilde{c}_{4}^1 \cdot 10^4$ & $2.37645$ & $-3.41042$ & $0.80760$ & $-0.12645$ & $2.64875$ & $6.83224$ \\
& $\tilde{c}_{5}^1 \cdot 10^4$ & $-3.55631$ & $4.16924$ & $-1.47993$ & $-0.90679$ & $-0.88886$ & $-1.90360$ \\
& $\tilde{c}_{6}^1 \cdot 10^4$ & $0.36494$ & $-1.41479$ & $1.04950$ & $0.01291$ & $2.09126$ & $4.01719$ \\
& $\tilde{c}_{7}^1 \cdot 10^5$ & $-1.34241$ & $6.21121$ & $-4.85371$ & $-0.01873$ & $-6.43171$ & $-12.48272$ \\
& $\tilde{c}_{8}^1 \cdot 10^5$ & $-5.29196$ & $3.08373$ & $1.19288$ & $-1.00350$ & $8.88013$ & $17.16267$ \\
& $s_R^1\,[\text{GeV}^2]$ & $0.58119$ & $0.01869$ & $-0.00078$ & $0.59911$ & $0.63227$ & $0.64046$ \\
\midrule
\multirow{9}{*}{\rotatebox{90}{$S2$-wave}} & $c_{0}^2$ & $-0.68903$ & $0.18537$ & $-0.03212$ & $-0.53549$ & $-0.32821$ & $-0.32336$ \\
& $c_{1}^2$ & $0.26966$ & $-0.12084$ & $0.03425$ & $0.18267$ & $0.09589$ & $0.09920$ \\
& $c_{2}^2 \cdot 10$ & $-0.84705$ & $0.60045$ & $-0.20249$ & $-0.44644$ & $-0.14089$ & $-0.13797$ \\
& $c_{3}^2 \cdot 10^2$ & $3.01653$ & $-2.84278$ & $1.13116$ & $1.28788$ & $0.30620$ & $0.31298$ \\
& $c_{4}^2 \cdot 10^2$ & $-1.31882$ & $1.49197$ & $-0.64268$ & $-0.45910$ & $-0.06999$ & $-0.06785$ \\
& $c_{5}^2 \cdot 10^3$ & $5.54428$ & $-7.29191$ & $3.39123$ & $1.58425$ & $0.18109$ & $0.18436$ \\
& $c_{6}^2 \cdot 10^3$ & $-2.45602$ & $3.48192$ & $-1.67013$ & $-0.61395$ & $-0.04298$ & $-0.03605$ \\
& $c_{7}^2 \cdot 10^3$ & $0.87971$ & $-1.35501$ & $0.67756$ & $0.18947$ & $0.01081$ & $0.00997$ \\
& $c_{8}^2 \cdot 10^4$ & $-2.69610$ & $3.93627$ & $-1.89379$ & $-0.61980$ & $-0.00785$ & $0.02123$ \\
\bottomrule
\end{tabular}
\caption{Chebyshev expansion coefficients and real-axis resonance
  parameters $s_R^I$ (where phase shifts pass through $\pi/2$) specifying
  the central Roy solutions as a function of the pion mass. The
  coefficients $c^I_{k,n}$ describe the quadratic pion-mass dependence,
  $c^I_k(y_\pi) = c^I_{k,0} + c^I_{k,1} y_\pi + c^I_{k,2} y_\pi^2$ with
  $y_\pi \equiv (M_\pi / M_{\pi,\text{phys}})^2$, which can be used both
  below and above the physical mass. For the $P$ wave, $\tilde{c}_k^1
  \equiv c_k^1 / (4 M_\pi^2)$.  For convenience, the exact values at the
  physical pion mass, $c^I_k(\text{phys})$, as well as for the central
  solution and the best fit to the lattice data at $M_\pi =
  239\,\text{MeV}$, $c^I_{k}(239\,\text{C})$ and $c^I_{k}(239\,\text{F})$,
  are also listed.}
\label{tab:chebyshev_coeffs}
\end{table}

\clearpage
\bibliographystyle{apsrev4-1_mod_2}
\bibliography{roychiral}

\end{document}